\documentclass[aps,floatfix,prb,reprint,notitle,
]{revtex4-2}

\usepackage{braket}
\usepackage{graphicx}
\usepackage{bm}
\usepackage{xcolor}
\usepackage{siunitx}
\usepackage{chemformula}
\usepackage{esint}

\newcommand{\eqname}{Eq.}
\newcommand{\secname}{Sec.}

\def\Xint#1{\mathchoice
   {\XXint\displaystyle\textstyle{#1}}%
   {\XXint\textstyle\scriptstyle{#1}}%
   {\XXint\scriptstyle\scriptscriptstyle{#1}}%
   {\XXint\scriptscriptstyle\scriptscriptstyle{#1}}%
   \!\int}
\def\XXint#1#2#3{{\setbox0=\hbox{$#1{#2#3}{\int}$}
     \vcenter{\hbox{$#2#3$}}\kern-.5\wd0}}

\def\dashint{\Xint-}

\newcommand{\Rcal}{\mathcal{R}}
\newcommand{\chis}{\stackrel{\text{\tiny(S)}}{\chi}}
\newcommand{\Rcals}{\stackrel{\text{\tiny(S)}}{\Rcal}}
\newcommand{\Rcalt}{\stackrel{\text{\tiny(T)}}{\Rcal}}
\newcommand{\Phis}{\stackrel{\text{\tiny(S)}}{\Phi}}
\newcommand{\Phit}{\stackrel{\text{\tiny(T)}}{\Phi}}
\newcommand{\Phithree}{\stackrel{\text{\tiny(3)}}{\Phi}}
\newcommand{\Phifour}{\stackrel{\text{\tiny(4)}}{\Phi}}
\newcommand{\bPhithree}{\boldsymbol{\Phithree}}
\newcommand{\bPhifour}{\boldsymbol{\Phifour}}

\newcommand{\bR}{\boldsymbol{R}}

\newcommand{\bGt}{\stackrel{\text{\tiny(T)}}{\boldsymbol{G}}}
\newcommand{\bGs}{\stackrel{\text{\tiny(S)}}{\boldsymbol{G}}}

\newcommand{\bPhi}{\boldsymbol{\Phi}}
\newcommand{\bPhis}{{\stackrel{\text{\tiny(S)}}{\bPhi}}}
\newcommand{\bPhit}{{\stackrel{\text{\tiny(T)}}{\bPhi}}}

\newcommand{\bUps}{\boldsymbol{\Upsilon}}

\newcommand{\bRcal}{\boldsymbol{\Rcal}}
\newcommand{\bRcalt}{{\stackrel{\text{\tiny(T)}}{\bRcal}}}
\newcommand{\bRcals}{{\stackrel{\text{\tiny(S)}}{\bRcal}}}

\newcommand{\bLambda}{\boldsymbol{\Lambda}}

\newcommand{\bp}{\boldsymbol{p}}

\renewcommand{\Im}{\mathrm{Im}}
\renewcommand{\Re}{\mathrm{Re}}

\renewcommand{\bf}{\boldsymbol{f}}

\newcommand{\br}{{\bm r}}

\newcommand{\Pis}{\stackrel{\text{\tiny(S)}}{\Pi}}
\newcommand{\Pit}{\stackrel{\text{\tiny(T)}}{\Pi}}
\newcommand{\bPis}{\stackrel{\text{\tiny(S)}}{\boldsymbol{\Pi}}}
\newcommand{\bPit}{\stackrel{\text{\tiny(T)}}{\boldsymbol{\Pi}}}

\newcommand{\bchi}{\boldsymbol{\chi}}
\newcommand{\chitwo}{\stackrel{\text{\tiny(2)}}{\chi}}
\newcommand{\bchitwo}{\mathbf{\stackrel{\text{\tiny(2)}}{\bchi}}}

\DeclareSIUnit\hartree{\protect \text {\protect \ensuremath {E}}_{\protect \mathrm {h}}}
\DeclareSIUnit\bohr{\protect \text {\protect \ensuremath {a}}_{0}}

\begin{document}

\preprint{APS/123-QED}

\title{Simulating the anharmonic phonon spectrum in critical systems:\\ self-consistent phonons and temperature-dependent effective potential methods}

\author{Lorenzo Monacelli}
 \affiliation{Dipartimento di Fisica, Universit\`a di Roma Sapienza}

\date{\today}

\begin{abstract}
Understanding and simulating the thermodynamic and dynamical properties of materials affected by strong ionic anharmonicity is a central challenge in material science. Much interest is in material displaying critical displacive behaviour, such as near a ferroelectric transition, charge-density waves, or in general displacive second-order transitions. In these cases, molecular dynamics suffer from a critical slowdown and emergent long-range fluctuations of the order parameter. Two prominent methods have emerged to solve this issue: Self-consistent renormalization of the phonons like the Self-Consistent Harmonic Approximation (SCHA) and Self-Consistent Phonons (SCP), and methods that fit the potential energy landscape from short molecular dynamics trajectories, like the Temperature-Dependent Effective Potential (TDEP). Despite their widespread use, the limitations of these methods are often overlooked in the proximity of critical points.

Here, we establish a guiding rule set for the accuracy of each method on critical quantities: free energy for computing the phase diagrams, static correlation functions for inferring phase stability and critical behaviours, and dynamic correlation functions for vibrational spectra and thermal transport. Also, a new TDEP implementation is introduced to fix the calculation of dynamical spectra, restoring the correct perturbative limit violated by the standard TDEP approach.

Results are benchmarked both against an exact one-dimensional anharmonic potential and two prototypical anharmonic crystals: the ferroelectric \ch{PbTe} and the metal-halide perovskite \ch{CsSnI3}.

\end{abstract}
\maketitle

\section{Introduction}
The study of critical phenomena in crystals near second-order phase transitions has shaped foundational theories in statistical mechanics for nearly a century. In displacive transitions, such as ferroelectricity and charge-density wave formation, the order parameter arises from symmetry-breaking atomic displacements. As the system approaches the critical point, these displacements drive anomalous lattice dynamics, marked by diverging nuclear oscillations. This crucial enhancement of anharmonic interactions leads to prolific phonon-phonon and electron-phonon scattering, which underpins functional properties in materials ranging from thermoelectricity, photovoltaics, superconductivity, and other correlated quantum systems.

The simulation of lattice dynamics near critical points presents several challenges. The excitation energy and dynamics of the order parameter go to zero, causing the divergence of the equilibration time. At the same time, the correlation length of the atomic fluctuations increases, and bigger and bigger simulation cells are required to converge the results. For this reason, molecular dynamics faces fundamental challenges, even when empirical or machine-learned force fields are available for evaluating the interatomic potential.


A particularly relevant case is the study of thermal transport. Here, phonon-phonon scattering is the major limiting factor for heat propagation as temperature increases, with profound technological implications for producing materials with extremely high values of thermal conductivity (heat dissipators) or particularly low ones (insulators).
In fact, materials with both low thermal conductivity and electrical resistivity are promising thermoelectrics that can extract energy from heat wasted by other processes.
Anharmonic phonon scattering is one of the best strategies to hamper heat diffusion in these good conductors, and the currently best-known bulk thermoelectric material, \ch{SnSe}, exploits a critical point in the ferroelectric phase-transition near the working temperature of \SI{800}{\kelvin}\cite{aseginolaza_phonon_2019}.

An exact approach to evaluate thermal transport in materials involves the calculation of the Green-Kubo heat-current autocorrelation function within equilibrium molecular dynamics\cite{Schelling2002} (MD) or, alternatively, performing non-equilibrium MD simulations with a sustained temperature gradient\cite{Tenenbaum1982}.
However, these methods are extremely computationally expensive to converge for the system size and simulation time, even when fast interatomic force fields are available\cite{Drigo2023}.
Approximate empirical methods based on the Boltzmann transport equation and its quantum counterpart, Wigner transport\cite{simoncelli_unified_2019}, emerged as the \emph{de facto} state-of-the-art to quantify the thermal conductivity.
However, these methods assume a perturbative treatment of phonon-phonon interaction. What happens when they are applied to strongly anharmonic materials, where the harmonic approximation and the quasiparticle picture of phonons break down? 

The same question arises with other properties depending on lattice motion, like the computation of vibrational spectroscopy within IR and Raman, related respectively to the dipole-dipole and polarizability-polarizability dynamical correlation functions, or the anharmonic phonon dispersion probed by Neutron and inelastic X-ray scattering, associated with the dynamical structure factor. 
In the presence of strong anharmonicity, such as in proximity with second-order displacive phase transitions, the susceptibility diverges, causing a significant deviation of the phonon energies from their harmonic ones, which perturbation theory cannot describe.

Many methodologies have been introduced to address these issues while maintaining the easy phonon quasiparticle description of lattice vibration. 
Since its first formulation in 1912\cite{Born1912}, the Self-Consistent Phonons (SCP) was successfully applied to describe the phonons of noble gasses in their solid phase\cite{Klein1972}. Inspired by SCP, many other methods attempted to bridge the concept of phonons to strongly anharmonic materials, employing a first-principles approach.
The Self-Consistent Ab Initio Lattice Dynamics (SCAILD)\cite{SCAILD,QSCALID} was one of the first methods applied in tandem with \emph{ab initio} simulations. Not too long after, many other methodologies emerged\cite{zhang_phonon_2014}, some commutated from quantum chemistry to periodic systems, like the vibrational self-consistent field \cite{monserrat_anharmonic_2013}. Among them, the most successful ones in terms of the number of applications and complexity of the materials analyzed are the stochastic self-consistent harmonic approximation (SCHA\cite{Errea2012,errea_anharmonic_2014,monacelli_stochastic_2021}) and its close relative, self-consistent phonons (SCP\cite{tadano_self-consistent_2015}). In this work, we will refer to SCHA to summarize different implementations of self-consistent harmonic theories.

A different route to compute anharmonic properties of solids consists of speeding up the convergence of molecular dynamics simulation, extracting from short trajectories observables that are fast to converge, and could then be related to slow-converging observables. The main approach here is the temperature-dependent effective potential (TDEP\cite{hellman_lattice_2011,hellman_temperature-dependent_2013,hellman_temperature-dependent_2013-1}), where an effective, temperature-dependent, polynomial potential is fitted from the energies and forces of a short MD trajectory and employed to solve the thermodynamics and dynamics of the system. Inspired by the success of the TDEP method, multiple other approaches that fit the potential energy landscape from MD trajectories have emerged\cite{bottin_-tdep_2020,Fransson2023,meitz_phonon_2025}, and we will refer to all of them as `TDEP' for simplicity.

All these methods correct the harmonic phonon dispersion with a nonperturbative self-energy, which can be exploited to compute thermodynamic quantities much more efficiently than standard MD. E.g., thermal conductivity may be obtained by adequately augmenting the Boltzmann-Wigner empirical transport equation\cite{simoncelli_unified_2019} with anharmonic spectral functions\cite{caldarelli_many-body_2022,dangic_origin_2021}. A similar approach has been recently applied to unveil the impact of anharmonicity on superconductivity in high-pressure hydrides\cite{Dangic2024}.
However, it has been shown that different theories lead to distinct phonon dispersions when applied in strongly anharmonic materials\cite{Fransson2023}. 

Are the aforementioned methods to simulate phonons in anharmonic systems an uncontrolled form of approximation with questionable predictive power, or is there a way to assess and systematically improve them, determining their accuracy \emph{a priori}?

\secname~\ref{sec:exact} presents the exact definition of phonons in a strongly anharmonic material. Then, \secname~\ref{sec:scha:tdep} introduces the equilibrium SCHA and TDEP methods from a rigorous variational principle on the free energy. In particular, TDEP is reformalized, underlying the only implementation based on proper first-principles grounds. \secname~\ref{sec:static:response} discusses the static linear-response theory applied to SCHA and TDEP, unveiling the origin of the observed discrepancies\cite{Fransson2023} and showing how they can be predicted \emph{a priori} and corrected within both theories.
The dynamical response theory is revised in \secname~\ref{sec:dynamical}, where the standard TDEP dynamical response function is shown to overcount anharmonicity, violating the perturbative limit. This work introduces a workaround to restore the proper anharmonic limit without extra computational cost. 
Finally, everything examined in this work is benchmarked against the exact solution of a simple anharmonic potential in \secname~\ref{sec:numerical}.
The similarities and differences between SCHA and SCP are discussed in \secname~\ref{sec:scp}. 
\secname~\ref{sec:conclusions} presents the perspectives of this work's findings. 
This work accounts for nuclei as classical particles for simplicity. However, some of our findings are general and also apply to quantum nuclei.

\section{Anharmonic phonons}
\label{sec:exact}

Phonons are dynamical excitations of the lattice. In a perfectly harmonic crystal, the energy of phonons coincides with the squared eigenvalues of the mass-reduced force constant matrix, and their vibrational pathway is a straight line identified by the eigenvectors. Conversely, anharmonicity bends the vibration pathway, which changes in time until the energy is dissipated throughout the whole system into incoherent vibrations in the thermal bath. The resulting finite lifetime of lattice vibrations in anharmonic crystals also introduces a broad energy width. This energy spectrum can be probed by looking at the absorption and resonances with external excitation (e.g., light). Therefore, the proper definition of phonons in anharmonic crystal is the energy spectrum of the lattice: the dynamic response function to an external perturbation.

The exact classical probability distribution of ions in phase space is defined as
\begin{equation}
    \rho_0(\bm p, \bm r) =
    \frac{1}{Z} \exp\left[-\beta H(\bm p, \bm r)\right],
    \label{eq:rho:exact}
\end{equation}
where $\bm r$ and $\bm p$ are the vectors of position and momentum of each particle in the system, $\beta$ the Boltzmann factor $\beta = 1 / k_bT$, $H(\bm p,r)$ is the Hamiltonian of nuclei (within the Born-Oppenheimer approximation) and $Z$ is the partition function defined as a normalization factor for the probability density $\rho$
\begin{equation}
\label{eq:part:func}
Z = \int d^{3N}p\, d^{3N}r \exp\left[-\beta H(\bm p, \bm r)\right].
\end{equation}
The knowledge of the partition function allows for computing equilibrium properties of matter through the free energy. At constant temperature, volume, and number of particles (canonical ensemble), the free energy is the one defined by Helmholtz as 
\begin{equation}
    F(N, V, T) = -\frac1\beta \ln Z.
\end{equation}
All thermodynamics observables can be expressed as derivatives of $F$ to $N, V, T$. In particular, the second derivatives of $F$ are static linear-response quantities discussed in \secname~\ref{sec:static:response}.

The generalization of phonons in anharmonic crystals are defined through the lattice's dynamic response to an external time-dependent perturbation on the nuclei, which coincides with the spectral features excited through a spectroscopy measurement.
The dynamical response is governed by the time propagation of the density distribution by the Liouville equation, acting as a master equation of the system:
\begin{equation}
\frac{\partial\rho}{\partial t} = \sum_{i = 1}^{3N}\left( \frac{\partial H}{\partial r_i}\frac{\partial \rho}{\partial p_i} - \frac{\partial H}{\partial p_i}\frac{\partial\rho}{\partial r_i}\right).
\label{eq:liouville}
\end{equation}
$\rho(\bp, \br, t_0)$ is prepared in equilibrium at time $t\le t_0$ where the density distribution satisfies \eqname~\eqref{eq:rho:exact}, which is a stationary solution of the master equation (\eqname~\ref{eq:liouville}).
At $t=t_0$, the system interacts with an external perturbation $H'(t)$, and the density matrix evolves. In a typical experiment, this perturbation is given by external electric fields (like in the Raman or Infrared), X-rays, or neutron beams. In numerical simulations, perturbing the system with a Dirac delta $\delta(t - t_0)$ impulse is convenient as it excites all frequencies with the same intensity. This is equivalent to offsetting the equilibrium distribution by a finite momentum along the perturbation direction and studying the free dynamics of the perturbed system.
The phonon Green function $G_{ij}(t)$ is defined as the dynamical response function of the average position of a nucleus when another nucleus is perturbed by a time-dependent external force
\begin{equation}
    H'(t) = r_j F(t)
\end{equation}
\begin{equation}
    \braket{r_i}(t) = \int_{t_0}^t G_{ij}(t - t') F(t') dt'
\end{equation}
where 
\begin{equation}
    \braket{r_i}(t) = \int d^{3N}r\, d^{3N}p\; r_i \rho(\bm r, \bm p, t)
    \label{eq:avgpos}
\end{equation}
and $F(t)$ is the time profile of the perturbation force on the atom $j$.
Most of the physical observables can be expressed as functions of the phononic Green's function. Of particular relevance is the so-called spectral function $\sigma(\omega, \bm q)$, which refers to the Fourier transform (both in time and space) of the phononic Green function:
\begin{equation}
    \tilde G_{ij}(\omega, \bm q) = \frac{1}{N_q}\sum_{j}e^{i \bm q \cdot (\bm r_i -  \bm r_j) - i \omega t} G_{ij}(t),
    \label{eq:green}
\end{equation}
where the sum over $j$ only runs on the periodic images of the $i$ with respect to the primitive cell, and $N_q$ counts the number of supercells included in the summation.
The spectral function is the trace of the imaginary part of the Green function
\begin{equation}
    \sigma(\omega, \bm q) = - \sum_{i = 1}^{3N_\text{uc}} \Im \tilde G_{ii}(\omega, \bm q)
    \label{eq:sigma},
\end{equation}
where $N_\text{uc}$ counts the number of atoms in the primitive cell.
The spectral function $\sigma(\omega, \bm q)$ presents peaks corresponding with the phonon excitation energies and is proportional to the response function of dynamical experiments, like Raman and IR spectroscopy, X-Ray, and Neutron scattering. The peaks of the spectral function are the real lattice excitations in the presence of anharmonicity and the natural extension of phonons in anharmonic crystals.
These results remain valid in the quantum regime as well, where $\rho(\bp, \br)$ becomes the density matrix, the Hamiltonian is an operator, and \eqname~\eqref{eq:liouville} must be replaced by the Liouville–von Neumann equation.

\section{Thermodynamics and phase-transitions within SCHA and TDEP}
\label{sec:scha:tdep}

The exact resummation of the partition function in \eqname~\eqref{eq:part:func} quickly becomes intractable for a system containing more than a few degrees of freedom. While there are numerical tools like Monte Carlo and Molecular Dynamics simulations to sample the equilibrium probability distribution $\rho_0(\bp, \br)$, the complete partition function is required to simulate the free energy (\eqname~\ref{eq:free:def}). 
This is where methods like SCHA and TDEP come in handy in providing an approximate definition of the free energy, which can be numerically computed very efficiently even with a system with tens of thousands of atoms.
Both methods approach the problem of computing the free energy by introducing a Gaussian trial density matrix defined as
\begin{equation}
    \tilde\rho(\bm p, \bm r) = \frac{ e^{ -\frac{\beta}{2} \left(\sum_{i =1 }^{3N} \frac{p_i^2} {m_i} +
    \sum_{ij = 1}^N \Phi_{ij}(r_i - \Rcal_i)(r_j - \Rcal_j)\right)}}{\mathcal Z},
    \label{eq:rho:scha}
\end{equation}
\begin{equation}
    \mathcal Z = \frac{\left(2\pi k_BT\right)^{3N}}{\sqrt{\det\bPhi}}\prod_{i = 1}^{3N} \sqrt{m_i}.
\end{equation}
This density matrix is the equilibrium solution of an auxiliary harmonic Hamiltonian
\begin{equation}
    \mathcal H(\bp, \br) = \sum_{i = 1}^{3N}\frac{p_i^2}{2m_i} + \frac 12 \sum_{ij}(r_i - \Rcal_i)\Phi_{ij}(r_j - \Rcal_j),
\end{equation}
\begin{equation}
    \mathcal V(\br) = \frac 12 \sum_{ij}(r_i - \Rcal_i)\Phi_{ij}(r_j - \Rcal_j).
\end{equation}
The Gaussian density matrix depends on two tensorial parameters: the average position $\bRcal$ and the auxiliary force constant matrix $\bPhi$.
Regardless of the trial density matrix, the exact free energy can be evaluated through thermodynamic integration, smoothly changing the probability distribution from the auxiliary Harmonic one into the real one 
\begin{equation}
    H(\lambda) = \sum_{i = 1}^{3N}\frac{p_i^2}{2m_i} + \mathcal V(\br) + \lambda \left[V(\br) - \mathcal V(\br)\right].
\end{equation}
When $\lambda=0$, we have the auxiliary harmonic Hamiltonian, with the probability distribution $\tilde\rho(\bm p, \br)$ defined in \eqname~\eqref{eq:rho:scha}. When $\lambda=1$, the Hamiltonian becomes the exact one, and the probability distribution is given by \eqname~\eqref{eq:rho:exact}.
The exact free energy can be evaluated by integrating the free energy derivative with respect to $\lambda$:
\begin{equation}
    F = F(\lambda\rightarrow 1) =  F(\lambda\rightarrow 0) + \int_0^1 \frac{\partial F}{\partial \lambda} d\lambda,
    \label{eq:thermodynamic:int}
\end{equation}
\begin{equation}
    \frac{\partial F}{\partial \lambda} = \left<\frac{\partial H}{\partial\lambda}\right>_\lambda =
    \braket{V - \mathcal V}_\lambda,
\end{equation}
\eqname~\eqref{eq:thermodynamic:int} is in principle exact. However, it requires many different simulations to sample the probability distribution $\rho(\boldsymbol{p}, \br, \lambda)$, which is often impractical.
SCHA and TDEP offer approximate solutions to the integral in \eqname~\eqref{eq:thermodynamic:int}.
In particular, it is possible to prove that the integrand $\braket{V-\mathcal V}$ is always a strictly decreasing function in $\lambda$ between 0 and 1. Therefore
\begin{equation}
    \braket{V-\mathcal V}_{\lambda=0} \ge \int_{0}^1\braket{V - \mathcal V}_\lambda d\lambda \ge \braket{V - \mathcal V}_{\lambda=1}.
    \label{eq:variational}
\end{equation}
The SCHA and TDEP exploit this property using two different strategies. 
From \eqname~\eqref{eq:variational} we can
derive a variational expression for the upper and lower bound of the free energy
\begin{equation}
    F_0 + \braket{V - \mathcal V}_{\lambda=0} \ge
    F \ge F_0 + \braket{V - \mathcal V}_{\lambda=1}.
    \label{eq:full:var}
\end{equation}
The SCHA exploits the left side of the inequality \eqname~\eqref{eq:full:var}, while TDEP the right side.
\begin{equation}
    F \le F_\bPhis + \braket{V(\br) - \mathcal V_{\bRcal,\bPhis}(\br) }_{\bRcals,\bPhis}
    \label{eq:var:scha}
\end{equation}
\begin{equation}
    F \ge F_\bPhit + \braket{V(\br) - \mathcal V_{\bRcalt,\bPhit}(\br) }_\text{MD}
    \label{eq:var:tdep}
\end{equation}
The difference between the two expressions relies on the ensemble on which the average on the right-hand side is evaluated. For \eqname~\eqref{eq:var:scha}, the ensemble is distributed according to the Gaussian in \eqname~\eqref{eq:rho:scha}. The average in \eqname~\eqref{eq:var:tdep}, instead, is evaluated when $\lambda=1$, thus on the exact distribution which has to be sampled using molecular dynamics (MD) or Monte Carlo techniques (which is referred to as a $\braket{\cdot}_\text{MD}$ in this work).
Both inequalities hold for any auxiliary Harmonic Hamiltonian. Therefore, SCHA and TDEP select the dynamical matrix to obtain the best possible free energy estimation. For the SCHA, this corresponds to minimizing the right-hand side of \eqname~\eqref{eq:var:scha}
\begin{equation}
    F_{\text{SCHA}} = \min_{\bPhis, \bRcals}F_\bPhis + \braket{V(\br) - \mathcal V_{\bRcal,\bPhis}(\br) }_{\bRcals,\bPhis}.
    \label{eq:free:scha}
\end{equation}
The condition that cancel the gradient of $F(\bPhis, \bRcals)$ give rise to two self-consistent equations\cite{monacelli_stochastic_2021}:
\begin{equation}
    \Phis_{ab} = \left<\frac{\partial^2 V}{\partial r_a\partial r_b}\right>_{\bRcals,\bPhis}
    \label{eq:phi:scha}
\end{equation}
\begin{equation}
    \left<\frac{\partial V}{\partial r_a}\right>_{\bRcals,\bPhis} = 0.
    \label{eq:rc:scha}
\end{equation}
Analogously, the TDEP auxiliary dynamical matrix and centroids maximize  the right-hand side of \eqname~\eqref{eq:var:tdep}
\begin{equation}
F_\text{TDEP} = \max_{\bRcalt, \bPhit}F_\bPhit + \braket{V(\br) - \mathcal V_{\bRcalt,\bPhit}(\br) }_\text{MD}.
\label{eq:tdep:start}
\end{equation}
The zero gradient condition of \eqname~\eqref{eq:tdep:start} provides an equivalent expression for TDEP quantities:
\begin{equation}
    \bRcalt = \braket{\br}_\text{MD},
    \label{eq:rc:tdep}
\end{equation}
\begin{equation}
\left(\bPhit^{-1}\right)_{ab} = \frac{1}{k_bT}\left<
    (r_a - \Rcalt_a)(r_b - \Rcalt_b)
    \right>_\text{MD}.
    \label{eq:phi:tdep}
\end{equation}
Differently for the self-consistent SCHA equations (\eqname~\ref{eq:rc:scha} and \ref{eq:phi:scha}), 
\eqname~\eqref{eq:rc:tdep} and \eqname~\eqref{eq:phi:tdep} are one-shot: the right-hand side does not depend on the left-hand side and can be evaluated with one iteration. This comes at the cost of properly sampling the average position and displacement-displacement correlation function exactly within MD or MC. The proof of \eqname~\eqref{eq:rc:tdep} and \eqname~\eqref{eq:phi:tdep} is reported in \appendixname~\ref{app:tdep:proof}.
Evaluating \eqname~\eqref{eq:phi:tdep} is challenging near critical points; in fact, the position-position correlator is a quantity that diverges and may require a long-time trajectory to be adequately sampled via MD.
For this reason, the TDEP method exploits a clever approach to replace \eqname~\eqref{eq:phi:tdep} with a procedure that requires a short-time trajectory to converge and smaller cells\cite{hellman_lattice_2011,hellman_temperature_2013,hellman_temperature-dependent_2013-1}. 
In fact, $\bPhit$ can be directly computed by fitting a quadratic force fields on the real forces extracted from configurations sampled on an equilibrium MD trajectory.  \appendixname~\ref{app:tdep:proof:fit} demonstrates that \eqname~\eqref{eq:rc:tdep} and \eqname~\eqref{eq:phi:tdep} also minimize the following least-square cost function:
\begin{equation}
    \mathcal L = \sum_{a = 1}^{3N}\left<\left[f_a(\br) + \sum_b \Phit_{ab} (r_b - \Rcalt_b)\right]^2\right>_\text{MD} ,
    \label{eq:least:squares}
\end{equation}
\begin{equation}
\frac{\partial \mathcal L}{\partial \bRcalt} = 0
\qquad
    \frac{\partial\mathcal L}{\partial \bPhit} = 0
    \label{eq:tdep:fit}
\end{equation}
where
\begin{equation}
    f_a(\br) = -\frac{\partial V}{\partial r_a}.
\end{equation}
Therefore, the force constants matrix fitting the first-principle forces $\bm f$ sampled by MD or MC also maximizes the TDEP free energy. Interestingly, the order in which the average position $\bRcalt$ and the auxiliary force constant $\bPhit$ are fitted is not important, as the zero-gradient condition of $\bRcalt$ in  \eqname~\eqref{eq:tdep:fit} does not depend on $\bPhit$.
This equivalence provides formal grounds for the standard TDEP procedure while imposing some constraints. 
For example, the fit cost function must be a least-square (\eqname~\ref{eq:least:squares}), and the variational principle remains valid only for a linear fit. Both these choices are fundamental to preserve the variational principle, and their violation leads to wrong results.
E.g., fitting higher-order force constants simultaneously with $\bPhit$ in  \eqname~\eqref{eq:least:squares}, as suggested in the original TDEP works to improve the accuracy\cite{hellman_temperature-dependent_2013-1}, has the opposite detrimental effect of underplaying the role of anharmonicity: an infinite series of high-order force constant matrix recovers the Taylor expansion of the potential where the fitted quantities converge to the trivial terms of the Taylor series. A practical example is reported in \secname~\ref{sec:numerical}, where the fit up to the third order leads to much worse results for free energy and linear response calculations in all tested cases.
Fitting higher-order force constants from MD trajectories only makes sense to train cheaper and reliable potentials on which longer simulations can be performed, as implemented in ALAMODE\cite{tadano_anharmonic_2014}.
The other important limitation is the choice of the fitting procedure, which is not arbitrary. For example, if the least-square cost function (\eqname~\ref{eq:least:squares}) is replaced with any other cost function (like \eqname~3 from ref.\cite{hellman_lattice_2011}), the resulting final $\bPhit$ does not maximize the free energy, leading, by definition, to a worse evaluation of the free energy and thermodynamics properties.
In other words, it is essential to remember that, despite the name of TDEP, the fitting procedure to derive $\Rcalt$ and $\bPhit$ is just a numerical trick to speed up the convergence of \eqname~\eqref{eq:rc:tdep} and \eqname~\eqref{eq:phi:tdep}, not a procedure to construct a temperature dependent effective potential.

Both SCHA and TDEP can be extended to work in the quantum regime. In particular, for the SCHA, this comes at no extra cost, as the density matrix of a quantum harmonic oscillator remains analytical\cite{monacelli_stochastic_2021}. In contrast, it presents more difficulties for the TDEP, as molecular dynamics sampling must be replaced by quantum sampling by employing more complex techniques like Path-Integral Molecular Dynamics\cite{castellano_mode-coupling_2023}, which is considerably more expensive. 

Notably, both SCHA and TDEP account entirely for the \emph{mode-mixing}, i.e., the eigenvectors of the auxiliary force constant matrices ($\bPhis$ and $\bPhit$) are allowed to change during the optimization. This is not the case for some implementations of SCP, where the eigenvectors of $\bPhis$ are constrained to the Harmonic one to speed up the calculation.

\section{Static Response theory}
\label{sec:static:response}
Extending the definition of phonons from the harmonic theory to anharmonic systems at finite temperatures has attracted significant efforts, as it is crucial for addressing changes in phase stability upon heating and cooling. At \SI{0}{\kelvin}, neglecting quantum zero-point motion, the (meta)stability of a structure can be evaluated from the absence of imaginary phonons in the harmonic dispersion.
In fact, a phonon with an imaginary frequency corresponds to a negative curvature of the total energy along the atomic displacement oriented with the phonon polarization. Since even an infinitesimal displacement in that direction leads to an energy decrease, the structure is in a saddle point of the Born-Oppenheimer energy landscape and is, therefore, unstable.
Effects like strain, doping, or temperature can stabilize unstable configurations.
However, while it is easy to evaluate changes in the harmonic phonons with the volume, as captured by the quasi-harmonic approximation (QHA), describing changes in the phonon frequencies with the temperature at constant volume is much more challenging.

It is, therefore, tempting to interpret the eigenvalues of the auxiliary dynamical matrix provided by the SCHA and TDEP as temperature-dependent phonons to assess the stability of an atomic structure.
This has been done in many works without a proper theoretical justification\cite{hellman_temperature_2013,tadano_anharmonic_2014,errea_anharmonic_2014,tadano_self-consistent_2015,Borinaga2016,Borinaga2017,dangic_origin_2021,Fransson2023}; however,
looking for sign changes in eigenvalues of $\bPhis$ and $\bPhit$ is trivially wrong, as it is easy to prove that both $\bPhis$ and $\bPhit$ are positive definite and, when properly converged, never display negative eigenvalues, even when the selected structure is unstable. 
When evaluating the lowest order perturbative correction of anharmonicity to the harmonic phonons, it was clear that the SCHA auxiliary force constants $\bPhis$ miss a negative definite term.
Therefore, a common \emph{a posteriori} correction employed to the SCHA (and SCP) was to add this extra term empirically, the so-called \emph{bubble} diagram, to the auxiliary phonons, allowing some of the $\bPhis$ eigenvalues to become negative. This term was introduced in the original SCP theory for computing the phonon spectrum of solid Helium and Neon\cite{Klein1972,Goldman1970}, and it has been formalized in the context of the SCHA by Bianco et al. \cite{bianco_second-order_2017}, proving that it emerges naturally from the calculation of the finite temperature free energy Hessian.
Due to the positive definite nature of the TDEP force-constant matrix as well (which is evident in \eqname~\ref{eq:phi:tdep}), some works attempted to add the bubble diagram also to the phonon dispersion calculated from $\bPhit$\cite{dangic_origin_2021}.
However, this section shows that TDEP and SCHA strongly differ in this aspect, as the \emph{bubble} diagram is already accounted for within the TDEP auxiliary matrix, and adding it on top results in overcounting the anharmonicity.

The ionic static response function measures how much the average position of an atom changes when a static force is applied to it.
Applying a static force on atoms results in changing the Hamiltonian into
\begin{equation}
    H'(\bm \xi) = H - \sum_{b = 1}^{3N}\xi_b (r_b - \Rcalt_b)
\end{equation}
where $\xi_b$ is the force on the $b$ atom (including the Cartesian coordinate). The (static) response function $\chi_{ab}$ is defined as the derivative of the average atomic position with respect to the force applied to the system
\begin{equation}
    \chi_{ab} = \frac{\partial \left<r_b\right>_{\bm \xi}}{\partial \xi_a}.
    \label{eq:response:function}
\end{equation}
In the TDEP case, whose averages are performed on the exact equilibrium ensemble, the fluctuation-dissipation theorem holds (see \appendixname~\ref{app:fluctuation:dissipation}), leading to
\begin{equation}
    \frac{1}{k_bT}\left<
    (r_a - \Rcalt_a)(r_b - \Rcalt_b)
    \right>_\text{MD} = \left.\frac{\partial \braket{r_a - \Rcalt_a}_{H'(\bm\xi)}}{\partial \xi_b}\right|_{\bm \xi = 0}.
\end{equation}
Thanks to the equivalence in \eqname~\eqref{eq:phi:tdep}, the ionic static response function is directly related to the TDEP auxiliary force constant matrix
\begin{equation}
    \left(\bPhit^{-1}\right)_{ab} = \chi_{ab}.
    \label{eq:static:response:tdep}
\end{equation}
Notably, \eqname~\eqref{eq:static:response:tdep} is exact. Therefore, the TDEP auxiliary phonons can be rigorously interpreted as the exact (dressed up to any order) static response function.
The same is not true for the SCHA. In particular, the average position in \eqname~\eqref{eq:response:function} is evaluated on the self-consistent Gaussian distribution, and not on the exact one; therefore, when computing the derivative, an extra term appears due to the explicit dependence of $\bRcals$ and $\bPhis$ on $\bm \xi$.
In particular, the SCHA response function is defined by substituting the exact average with the one evaluated on the SCHA Gaussian distribution in \eqname~\eqref{eq:response:function}
\begin{equation}
    \chis_{ab} = \frac{\partial \left<r_a\right>_{\bRcals,\bPhis,\bm\xi}}{\partial\xi_b} = \frac{\partial\Rcals_a(\bm \xi)}{\partial\xi_b} .
    \label{eq:response:scha}
\end{equation}
To solve \eqname~\eqref{eq:response:scha}, we need to apply perturbation theory to the SCHA equations, similarly to perturbation theory is derived in the framework of other self-consistent theories like Hartree-Fock or DFT.
Differentiating the gradients of the SCHA equation with respect to the perturbation $\boldsymbol{\xi}$, we get
\begin{equation}
    \left<f_a\right>_{\Rcals(\xi),\Phis(\xi)} = \xi_a,
\end{equation}
\begin{equation}
    \frac{\partial}{\partial\xi_b}\left<f_a\right>_{\Rcals(\xi),\Phis(\xi)} = \delta_{ab};
\end{equation}
\begin{equation}
    \Phis_{ab}(\bm\xi) = \left<\frac{d^2V}{dR_adR_b}\right>_{\Rcals(\xi),\Phis(\xi)},
\end{equation}
\begin{equation}
    \frac{\partial\Phis_{ab}}{\partial \xi_c} = \frac{\partial}{\partial\xi_c} \left<\frac{d^2V}{dR_adR_b}\right>_{\Rcals,\Phis}.
\end{equation}
By following all mathematical trivial steps (reported in \appendixname~\ref{app:linear:response:scha}), one arrives at the two coupled linear response equations for the SCHA:
\begin{equation}
    \delta_{ab} =  -\sum_{h}\Phis_{ah} \frac{\partial\Rcals_h}{\partial\xi_b}  + \sum_{pqlm}\Lambda_{pqlm}\Phithree_{pqa}\frac{\partial\Phis_{lm}}{\partial\xi_b}, \label{eq:linear:response:rc}
\end{equation}
\begin{equation}
    \frac{\partial\Phis_{ab}}{\partial\xi_c} =  -\sum_{h}\Phithree_{abh} \frac{\partial\Rcals_h}{\partial\xi_c}  -\sum_{pqlm}\Lambda_{pqlm}\Phifour_{pqab}\frac{\partial\Phis_{lm}}{\partial\xi_c} ,
    \label{eq:linear:response:phi}
\end{equation}
where the 4-rank tensor $\bLambda$ is defined in \eqname~\eqref{eq:app:Lambda}, while $\bPhithree$ and $\bPhifour$ are the 3 and 4-phonon scattering vertices averaged on the ensemble
\begin{equation}
    \Phithree_{abc} = 
    \left<\frac{d^3V}{dR_adR_bdR_c}\right>_{\Rcals,\Phis},
    \label{eq:phi3}
\end{equation}
\begin{equation}
    \Phifour_{abcd} = 
    \left<\frac{d^4V}{dR_adR_bdR_cdR_d}\right>_{\Rcals,\Phis}.
    \label{eq:phi4}
\end{equation}
Inverting simultaneously \eqname~\eqref{eq:linear:response:rc} and \eqname~\eqref{eq:linear:response:phi} allows one to compute the SCHA static response function $\chis$ of \eqname ~\eqref{eq:response:scha}, derived formally for the first time in ref.\cite{bianco_second-order_2017}. This expression leads to defining a static self-energy $\Pi_{ab}$ that transforms the phonons from the auxiliary force constant matrix $\bPhis$ into the poles of the static response function as
\begin{equation}
    \chis_{ab} = \left[\Phis_{ab} + \Pi_{ab}\right]^{-1}
    \label{eq:static:response:scha}
\end{equation}
\begin{equation}
    \Pis_{ab} = \sum_{cdefgh}\Phithree_{acd}\Lambda_{cdef}\left[\mathbf{1} - \bPhifour\bLambda\right]^{-1}_{efgh}\Phithree_{ghb}
    \label{eq:self:energy}
\end{equation}
where $\mathbf{1}$ is the identity matrix and all the expressions between the square brakets are intended as matrix operations where each 4-rank tensors is treated as a matrix joining the first two and last two indices ($\Lambda_{abcd} \equiv \Lambda_{(ab)(cd)}$). This derivation is detailed in ref.\cite{bianco_second-order_2017}.  

The SCHA expression for the response function (\eqname~\ref{eq:static:response:scha}) is more complex than the TDEP one (\eqname~\ref{eq:static:response:tdep}) as it involves correcting the phonons coming from the free energy optimization through a self-energy. In particular, the auxiliary force constant matrix of the SCHA (\eqname~\ref{eq:phi:scha}) coincides with the inverse of the response function like the TDEP one only in the case that the self-energy $\bm\Pi$ vanishes, i.e., when $\bPhithree$ is precisely zero. Notably, no symmetry condition can enforce $\bPhithree$ to be zero, as the summation in \eqname~\eqref{eq:linear:response:rc} runs over the whole Brillouin zone. For example, the inversion symmetry, often wrongly invoked to assume a vanishing self-energy (\eqname~\ref{eq:self:energy}), cancels out only $\bPhithree$ matrix elements at $\Gamma$ across atoms mapped into themselves by the inversion symmetry. 
Interestingly, by considering $\bPhifour\to0$, the expression of the self-energy becomes much easier, and it coincides with the \emph{bubble} diagram derived from perturbation theory\cite{bianco_second-order_2017}. Indeed, the complete expression for the self-energy (\eqname~\ref{eq:self:energy}) goes beyond perturbation theory: expanding the matrix inversion is equivalent to solving a Dyson equation of RPA-like contributions, accounting for anharmonicity up to an arbitrary order\cite{bianco_second-order_2017}. 

The expression for the response function devised here and the SCHA free energy Hessian derived by Bianco et al.\cite{bianco_second-order_2017} are related as
\begin{equation}
\left(\boldsymbol{\chis}\right)^{-1}_{ab} = \frac{d^2 F_\text{SCHA}}{d\Rcals_a d\Rcals_b},
\end{equation}
where the dependence of the free energy landscape from the atomic position can be obtained by minimizing $F_\text{SCHA}$ only on the effective force constant matrix, constraining the centroids $\bRcals$. The free energy landscape defined in this way coincides with the Landau theory of phase transition: when the curvature of the free energy vanishes with respect to an order parameter $\Delta$ (function of the atomic positions), the corresponding response function $\boldsymbol{\chi}$ diverges.
This goes beyond the free energy landscape introduced by sampling transition rates in molecular dynamics, as the SCHA does not perform any adiabatic approximation on the order parameter's dynamics. 

The response functions of SCHA and TDEP have a key difference: the SCHA free energy Hessian can have imaginary frequencies, while the TDEP one can not. 
This arises from the symmetry constraints imposed on the SCHA density matrix during the free energy minimization. In fact, like in standard Harmonic phonons, if symmetries (either translational or point group) are accounted for, we can have a negative curvature of the free energy landscape along symmetry-breaking displacements. Conversely, the TDEP fit occurs on MD trajectories where no symmetries are imposed a priori, thus always leading to the thermodynamic stable structure, which has a positive definite free energy Hessian. In other words, we cannot use TDEP to study high-symmetry structures when they are unstable. This is a substantial advantage of the SCHA over TDEP for studying second-order phase transition, as they can be identified directly from the high-symmetry phase by looking for the crossing point when a phonon band changes from imaginary to real. 

Within the SCHA, we can identify a well-defined phase boundary in displacive transitions. 
The same is not true within the TDEP. The TDEP dynamical matrix is always positive definite (\eqname~\eqref{eq:phi:tdep}), and it coincides with the static response function (\eqname~\eqref{eq:static:response:tdep}). Moreover, since the two phases of a displacive transition continuously change one into the other, it is highly challenging to assign structures from the MD trajectories to one of the two phases, thus broadening the phase boundary. Indeed, the TDEP response function should diverge at the critical point and lead to a vanishing frequency in the TDEP dynamical matrix. However, this only occurs in the thermodynamic limit, as in any finite size system \eqname~\eqref{eq:phi:tdep} will lead to a broadened peak. This does not change if $\bPhit$ is obtained through the fitting procedure in \eqname~\eqref{eq:tdep:fit}, as its equivalence with \eqname~\eqref{eq:phi:tdep} holds at any given system size. 
Thus, converging to the thermodynamic limit in TDEP (or MD in general) is much trickier than in the SCHA, where a transition point exists even in finite-size systems due to the mean-field-like nature of the theory.

This analysis rationalizes the origin of the discrepancy between SCHA, SCP, and TDEP reported in ref.\cite{Fransson2023}, where the phonon frequencies of SCHA (and SCP) give an apparent worse agreement with the full MD calculations with respect to TDEP ones at low frequencies (thus close to the static limit). Indeed, the TDEP renormalized force constant matrix $\bPhit$ already coincides with the response function, while the SCHA (and SCP) require to invert the \eqname~\eqref{eq:linear:response:rc} and \eqref{eq:linear:response:phi} to obtain the phonons, as $\bPhis$ alone does not carry a valid physical meaning. Moreover, the case of metal halide perovskites deserves a special mention, as the fourth order $\bPhifour$ in these systems plays a significant role around the tetragonal-to-cubic phase transition, and the complete RPA-like resummation is needed to determine the phase stability. We show explicit calculation of this effect for \ch{CsSnI3} in \secname~\ref{sec:real:systems}.

A full description of the static susceptibility and the effect of different approximations in an exact 1D model is benchmarked in \secname~\ref{sec:numerical:static}.

\section{Dynamical response}
\label{sec:dynamical}

The dynamical nuclear response function probes how atoms respond to external time-dependent stimuli, such as optical electric fields in Raman and IR spectroscopies or X-rays and neutrons in inelastic scattering experiments. This quantity is essential for evaluating the spectral signatures of phonons. In particular, the peaks (or poles) of the spectrum derived from the dynamic susceptibility offer the most direct means of identifying anharmonic phonons, as they correspond to experimentally measurable observables.
Unlike the auxiliary force constant matrices, $\bPhit$ and $\bPhis$, the dynamical susceptibility can be directly accessed in experiments. For this reason, we refer to the poles of the nuclear dynamical response function as \emph{physical phonons} and to the square root of the eigenvalues of the mass-rescaled force constant matrices as \emph{auxiliary phonons}.
The nuclear spectral function is the dissipative (imaginary) part of the dynamic response function and can be measured in absorption experiments (e.g., IR spectroscopy). In strongly anharmonic crystals, it exhibits a broad spectrum, often featuring satellite peaks\cite{ribeiro_strong_2018} due to particular resonances of the phonon scattering events within the Brillouin zone. This spectral function plays a central role in determining phonon lifetimes and, consequently, the lattice thermal conductivity\cite{caldarelli_many-body_2022,dangic_lattice_2025}. Its importance becomes especially pronounced near critical points, where large fluctuations of the order parameter enhance phonon-phonon interactions.

\begin{figure}
    \centering
    \includegraphics[width=\columnwidth]{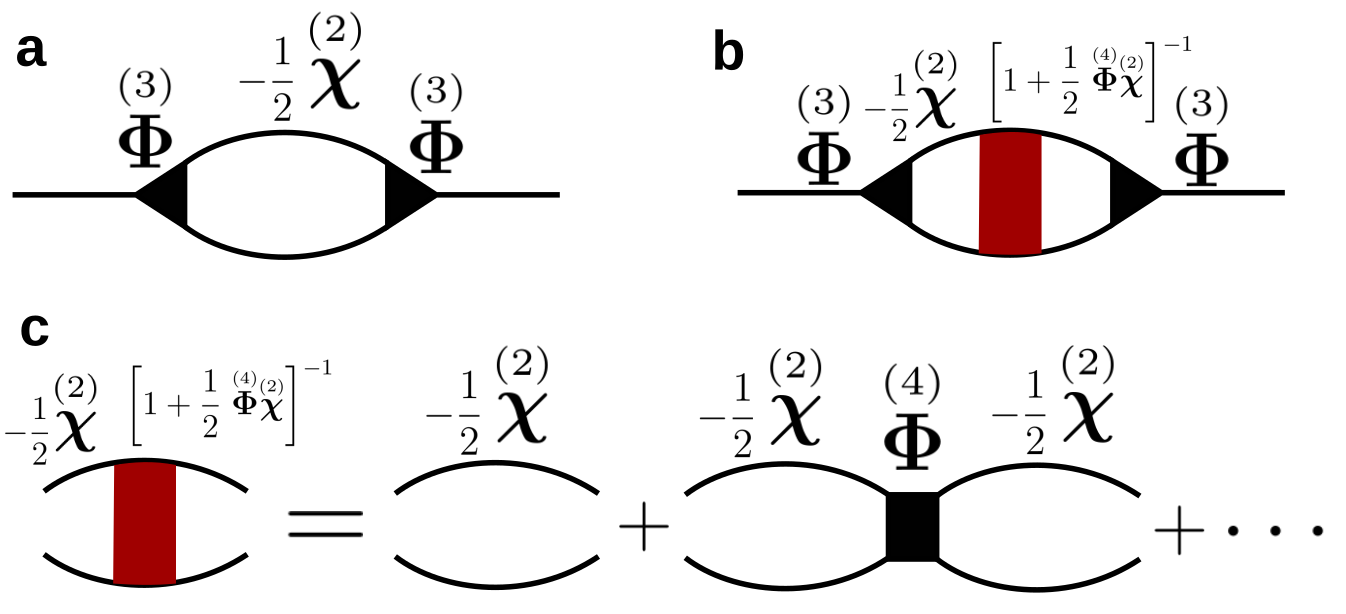}
    \caption{\small
    Diagrammatic representation of the TD-SCHA phonon self-energy (\eqname~\eqref{eq:dyn:self:energy}). \textbf{a} Self-energy in the \emph{bubble} approximation ($\bPhifour = 0$). \textbf{b} Full expression of the TD-SCHA self-energy (similar to RPA). \textbf{c} Diagrammatic expansion of the RPA phonon self-energy.\label{fig:self:energy}}
\end{figure}

Both TDEP and SCHA are static theories and, thus, can not calculate response functions to dynamic perturbation.
Recently, they have been extended in the time domain to address this issue. The SCHA evolved into the time-dependent self-consistent harmonic approximation (TD-SCHA), in which the density matrix is approximated as the most general time-dependent Gaussian. The parameters of the Gaussian evolve following the Dirac quantum least action principle\cite{lihm2020gaussian,monacelli_time-dependent_2021,siciliano_wigner_2023}. This allows the computation of the linear response to dynamic perturbations, which has been successfully applied to unveil the Raman/IR spectrum of high-pressure hydrogen and ice and the spectrum of metal halide perovskites\cite{monacelli_black_2020,Cherubini2021,monacelli_first-principles_2023}. Notably, the TD-SCHA put formal basis behind the dynamical \emph{ansatz} that was employed empirically: we can perform an analytic continuation of the static self-energy (\eqname~\ref{eq:self:energy}) in the complex plane $\bm \Pi \to \bm\Pi(z)$, then evaluating the self-energy at finite frequency $z \to \omega + i0^+$\cite{bianco_second-order_2017,Goldman1970, Klein1972}. This \emph{ansatz} has been proved simultaneously by two independent works to satisfy the variational dynamical principle\cite{lihm2020gaussian,monacelli_time-dependent_2021}, and it is equivalent to computing the dynamical response function directly within the TD-SCHA framework\cite{siciliano_wigner_2023}.
The analytic continuation is performed by introducing the two-phonons free propagator $\bchitwo(z)$
\begin{align}
\chitwo_{abcd}(z) = \sum_{\mu\nu}\frac{e_\mu^a e_\nu^b e_\mu^c e_\nu^d}{2\omega_\mu\omega_\nu}&\bigg[
\frac{(\omega_\mu + \omega_\nu)(1 + n_\mu + n_\nu)}{(\omega_\mu + \omega_\nu)^2 - z^2} -\nonumber \\ & + \frac{(\omega_\mu - \omega_\nu)(n_\mu-n_\nu)}{(\omega_\mu - \omega_\nu)^2 - z^2}
\bigg],
\end{align}
where $n_\mu$ is the Bose-Einstein occupation number for $\omega_\mu$. Noticing that
\begin{equation}
    \Lambda_{abcd} = -\lim_{z\to 0}\frac 12 \chitwo_{abcd}(z),
    \label{eq:static:limit}
\end{equation}
the analytic continuation of the self-energy is performed by substituting \eqname~\eqref{eq:static:limit} in \eqname~\eqref{eq:self:energy}, and removing the $z\to 0$ limit:
\begin{equation}
    \Pis_{ab}(z) = -\frac 12 \sum_{\substack{cdef\\gh}}\Phithree_{acd}\chitwo_{cdef}(z)\left[ \mathbf{1} + \frac 12 \bPhifour\bchitwo(z) \right]^{-1}_{efgh}\hspace{-0.3cm}\Phithree_{ghb},
    \label{eq:dyn:self:energy}
\end{equation}
from which it is possible to calculate the dynamical SCHA Green function:
\begin{equation}
    \bGs(z) = \left(z^2\mathbf{1} - \bPhis - \bPis(z)\right)^{-1}.
    \label{eq:chis}
\end{equation}
Notably, this expression accounts for quantum fluctuations thanks to the Bose-Einstein occupation (the SCHA is a quantum theory). 
The poles of \eqname~\eqref{eq:chis} are located at the square root of the eigenvalues of $\Phis + \Pis(z)$, where the real part of $\Pis(z)$ produces a shift of the phonon frequencies compared to the SCHA auxiliary phonons, while the imaginary part of $\Pi(z)$ broaden the spectrum, introducing a finite lifetime to phonons.
If one is familiar with perturbation theory, it is easy to recognize that, for $\bPhifour=0$,  \eqname~\eqref{eq:dyn:self:energy} resembles the self-energy of the \emph{bubble} diagram for phonon-phonon scattering (\figurename~\ref{fig:self:energy}\textbf{a}), with the key difference that the three-phonon scattering vertex is averaged on the SCHA ensemble (\eqname~\ref{eq:phi3}) and the free two-phonon propagator is evaluated with the converged auxiliary SCHA frequencies (eigenvalues of $\bPhis$) instead of harmonic phonons. For this reason, computing the self-energy in the \emph{bubble} approximation ($\bPhifour=0$) goes beyond perturbation theory, leading to extremely good results in multiple materials where perturbation theory fails, as demonstrated in the computation of \ch{SnSe} thermal conductivity near the critical phase transition\cite{aseginolaza_phonon_2019}.
Overcoming the \emph{bubble} approximation requires inverting the square brackets in \eqname~\eqref{eq:dyn:self:energy}, which leads to a RPA-like infinite series of contributions reported in \figurename~\ref{fig:self:energy}\textbf{c}. This task is exceptionally challenging, as one needs to invert a $N^2\times N^2$ matrix, where $N$ is the number of phonon modes in the whole Brillouin zone. Moreover, this inversion must be performed for each value of $z$. Recently, a more efficient Lanczos algorithm was devised to compute the full dynamical response function, which allowed the computation of dynamical spectra functions accounting for the complete RPA series shown in \figurename~\ref{fig:self:energy}\textbf{b,c}\cite{monacelli_time-dependent_2021}. This algorithm is analogous to the Lanczos approach employed in time-dependent density functional theory for electronic dynamical response functions to study excitonic effects and magnon spectra, e.g. TurboTDDFT\cite{Rocca2008}, TurboEELS\cite{timrov_2015}, and TurboMagnon\cite{gorni_2022}.

\vspace{0.3cm}
The first rigorous formalization of the dynamical response function for TDEP was introduced by Castellano \emph{et al}\cite{castellano_mode-coupling_2023} based on the mode-coupling theory. As for the SCHA, we can define a TDEP dynamical response function $\bGt(z)$ as
\begin{equation}
    \bGt(z) = \left(z^2\mathbf{1} - \bPhit - \bPit(z)\right)^{-1}.
    \label{eq:chit}
\end{equation}
However, the main challenge of extending the TDEP Green's function to finite frequency resides in defining the TDEP self-energy. In fact, in contrast with the SCHA, $\bPhit$ is already the exact inverse of the static response function: the TDEP self-energy must vanish for $z\to 0$. A simple analytical continuation of the self-energy leads to a trivial result
\begin{equation}
    \Pit_{ab}(z) = 0.
    \label{eq:pi:zero}
\end{equation}
Replacing \eqname~\eqref{eq:pi:zero} in \eqname~\eqref{eq:chit} unavoidably leads to a vanishing imaginary part and a nondissipative system, which prevents accessing crucial information like the phonon lifetime.
This problem can be manually ``fixed'' by computing the lifetime of phonons exploiting the Fermi golden rule by fitting the third-order force constant matrix together with $\bPhit$ in \eqname~\eqref{eq:least:squares}\cite{hellman_temperature_2013}. This strategy was employed by some TDEP implementations\cite{hellman_temperature-dependent_2013,bottin_-tdep_2020,xie_first_principles2020,li_first-principles_2023}. However, we already proved in \secname~\ref{sec:scha:tdep} that this fit violates the variational principle and must be avoided. An alternative strategy is to fit a third-order force constant on the residual forces $\delta\bf$, keeping $\bPhit$ fixed, similar to how high-order force constants are defined within the SCHA, where
\begin{equation}
    \delta f_a( \bR) = f_a(\bR) + \sum_b \Phit_{ab}(R_b - \Rcalt_b),
\end{equation}
\begin{equation}
    \min_{\bPhithree} \left<\hspace{-0.1cm}\sum_a\hspace{-0.1cm}\left(\hspace{-0.1cm}\delta f_a(\bR)\hspace{-0.08cm} + \hspace{-0.1cm} \frac 12\sum_{bc} \Phithree_{abc}(R_b - \Rcalt_b)(R_c - \Rcalt_c)\hspace{-0.1cm}\right)^{\hspace{-0.1cm}2}\right>_\text{MD} \hspace{-0.4cm} .
    \label{eq:tdep:fit:3}
\end{equation}
This approach is formally grounded in the mode-coupling theory\cite{castellano_mode-coupling_2023}. Once the third order $\bPhithree$ is fitted from the residual force minimizing the least squares in \eqname~\eqref{eq:tdep:fit:3}, TDEP exploits the Fermi golden rule to evaluate the imaginary part of the self-energy ($\Im\bPit$) within the \emph{bubble} approximation. This corresponds to computing the imaginary part of the SCHA dynamical self-energy $\Pis(z)$ within the \emph{bubble} approximation ($\bPhifour = 0$), where the free two-phonon propagator $\bchitwo(z)$ is evaluated with the TDEP phonon frequencies (square-root of the eigenvalues of the mass-rescaled $\bPhit$), and the three-phonons scattering vertex $\bPhithree$ is obtained from the fit of the residual forces $\delta \bf$ on the MD trajectory (\eqname~\ref{eq:tdep:fit:3}),
\begin{align}
    \Im\tilde \Pit_{\mu\mu}(\omega) =\frac{\pi}{16}\sum_{\nu\eta} \frac{\left|\tilde\Phithree_{\mu\nu\eta}\right|^2}{\omega_\mu\omega_\nu\omega_\eta}&\bigg[(n_\nu + n_\eta + 1)\delta(\omega - \omega_\nu - \omega_\eta) + \nonumber \\ & + 2(n_\nu - n_\eta)\delta(\omega - \omega_\nu + \omega_\eta)\bigg],
    \label{eq:fermi:rule}
\end{align}
where the $\tilde\bPhithree$ and $\tilde \Pit_{\mu\mu}$ have been transformed in the basis of eigenvectors of $\bPhit$:
\begin{equation}
    \tilde \Pit_{\mu\mu}(z) = \sum_{ab}\Pit_{ab}(z)\frac{e_\mu^a e_\mu^b}{\sqrt{m_a m_b}},
\end{equation}
\begin{equation}
    \tilde\Phithree_{\mu\nu\eta} = \sum_{abc}\Phithree_{abc}(z)\frac{e_\mu^a e_\nu^be_\eta^c}{\sqrt{m_a m_bm_c}},
\end{equation}
\begin{equation}
    \sum_{b} \frac{\Phit_{ab}e_\mu^b}{\sqrt{m_am_b}}= \omega_\mu^2 e_\mu^a
\end{equation}
This procedure leads to the imaginary part of the diagonal elements of the \emph{bubble} self-energy\cite{hellman_temperature-dependent_2013,castellano_mode-coupling_2023}. From \eqname~\eqref{eq:fermi:rule}, one can compute the real part of the diagonal elements of the self-energy $\Pit(\omega)$ using the Krames-Kroenig relations.
\begin{equation}
    \Re\Pit_{\mu\mu}(\omega) = \frac{1}{\pi}\dashint_{-\infty}^\infty d\omega' \frac{\Im \Pit_{\mu\mu}(\omega')}{\omega' - \omega},
    \label{eq:self:energy:tdep:real}
\end{equation}
where $\dashint$ indicates the Cauchy principal value of the integral. In principle, we have now all the ingredients to define the full dynamical spectral function within TDEP, by fitting the real and imaginary part of the self-energy (\eqname~\ref{eq:fermi:rule} and \ref{eq:self:energy:tdep:real}) inside the dynamical response function (\eqname~\ref{eq:chit}). From the interacting response function, one can obtain a new two-phonons density of states, and insert it back in \eqname~\eqref{eq:fermi:rule}, then recompute the full self-energy iteratively until convergence. This iterative approach corresponds to dressing the two-phonons propagators in \figurename~\ref{fig:self:energy}\textbf{a} with the interacting one-phonon Green's function. Notably, this dressing is not present in the SCHA theory.
In contrast to the static case, and despite the claim in Ref.~\cite{castellano_mode-coupling_2023}, this approach does not yield the exact dynamical results, as it relies on approximations that are systematically overcome within the SCHA framework. Specifically:
i) the phonon self-energy is truncated at the \emph{bubble} level (i.e., $\bPhifour = 0$), and
ii) only diagonal elements of the self-energy are considered, thereby neglecting the mode mixing introduced by the \emph{bubble} diagram (\figurename~\ref{fig:self:energy}\textbf{a}).
Although these approximations are often reasonable and are sometimes employed even on top of the SCHA, their validity must be assessed on a material-specific basis. For instance, the \emph{bubble} approximation breaks down in 2D materials\cite{Aseginolaza2024} and the cubic phase of metal halide perovskites\cite{monacelli_first-principles_2023}, producing qualitatively incorrect results. This breakdown is discussed in detail in \secname~\ref{sec:real:systems}.

Although this approach for the TDEP dynamical self-energy is, in principle, valid, it conceals a subtle but significant pitfall that requires careful attention. Specifically, the Kramers-Kronig relations used to compute the real part of the \emph{bubble} self-energy (\eqname~\ref{eq:self:energy:tdep:real}) do \emph{not} enforce the condition $\Pi(\omega \to 0) = 0$, which is required by the TDEP formalism. Instead, the Kramers-Kronig transformation relates the real and imaginary parts of an analytic function, defined only up to an arbitrary real constant. In \eqname~\eqref{eq:self:energy:tdep:real}, this constant is implicitly chosen so that $\Pi(\omega \to \infty) = 0$, which is appropriate for the SCHA self-energy (or the general \emph{bubble} diagram), but \emph{not} for TDEP. In TDEP, the correct condition would be $\Pi(\omega \to 0) = 0$. This discrepancy is evident from \eqname~\eqref{eq:self:energy:tdep:real} and \eqref{eq:fermi:rule}, where the shift is not adjusted to fulfill the TDEP constraint. In this way, the dynamical extension of TDEP violates the static limit and overcounts the real part of the \emph{bubble} diagram, already present inside $\bPhit$ in \eqname~\eqref{eq:chit}.

Luckily, this problem has a straightforward solution: removing the static limit of the \emph{bubble} self-energy from the calculation of the dynamical response function:
\begin{equation}
    \bGt(z) = \left[z^2\mathbf{1} - \bPhit - \bPit(z) + \bPit(z\to 0)\right]^{-1}
    \label{eq:dyn:tdep:fixed}
\end{equation}
\eqname~\eqref{eq:dyn:tdep:fixed} reproduces the correct $\omega\to 0$ limit and also satisfies the lowest-order perturbation theory at finite $\omega$. \eqname~\eqref{eq:dyn:tdep:fixed} also generates the correct second-order memory kernel for the mode-coupling theory derived in ref.\cite{castellano_mode-coupling_2023} while fixing the ambiguities of the Kramers-Kronig transformations.

Differently from the SCHA, where the dynamical Green's function is obtained from an RPA procedure, the correction in \eqname~\eqref{eq:dyn:tdep:fixed} is defined perturbatively, adding more scattering processes in the self-energy (or the memory kernel in the language of mode-coupling theory\cite{castellano_mode-coupling_2023}).
However, perturbative truncations of the self-energy, like the \emph{bubble} approximation, result in the violation of the response functions' sum rules, that are restored only by the complete RPA resummation \cite{Goldman1970,Klein1972,Aseginolaza2024}. This is crucial when computing properties for $q\to 0$, such as the bending rigidity of 2D materials\cite{Aseginolaza2024}, which may diverge if the self-energy is truncated at any finite order.

A numerical test for the perturbative regime of TDEP and SCHA dynamical spectral functions is reported in \secname~\ref{sec:test:dynamical}.

\section{Numerical comparison}
\label{sec:numerical}
This section tests the SCHA and TDEP thermodynamics (free energy), static response, and dynamical response functions against an exact (numerical) benchmark.

Let us consider the vibrational motion of a biatomic molecule in an anharmonic potential of the form
\begin{equation}
    V(x) = \frac 12 a x^2 + \frac{1}{6} b x^3 + \frac{1}{24} c x^4,
    \label{eq:1dpot}
\end{equation}
where $x$ represents the bond length.
The anharmonicity of this simple potential can be gradually switched on by tuning the $b$ and $c$ parameters.
To isolate the role played by the third-order anharmonicity (the $b$ parameter), we set $a=\SI{1.0}{\hartree\per\bohr^2}$ and $c=\SI{1.0e-4}{\hartree\per\bohr^4}$, just big enough to avoid runaway solutions while tuning $b$.

Although this may seem trivial, a systematic benchmark comparing TDEP and SCHA was still missing. 

To distinguish between different TDEP implementations, we will refer to \emph{step-by-step} as the procedure of fitting first a quadratic potential, and then third and higher-order force constants from the residual forces (\eqname~\ref{eq:tdep:fit:3}). Instead, we refer to the procedure of fitting all force constants simultaneously as TDEP \emph{one-shot}. The label ``+ bubble'' means that the static self-energy in the \emph{bubble} approximation is added on top of $\bPhit$ (\eqname~\ref{eq:fermi:rule}, \ref{eq:self:energy:tdep:real}).


\subsection{Free energy}

All thermodynamic quantities can be inferred from derivatives of the free energy. In this section, the free energy obtained within SCHA (\eqname~\ref{eq:free:scha}) and TDEP (\eqname~\ref{eq:tdep:start}) is benchmarked.  
To provide a comprehensive comparison, also the pure vibrational contribution to the TDEP free energy is evaluated, as performed by some TDEP implementations\cite{bottin_-tdep_2020}.

\begin{figure}
    \centering
    \includegraphics[width=\columnwidth]{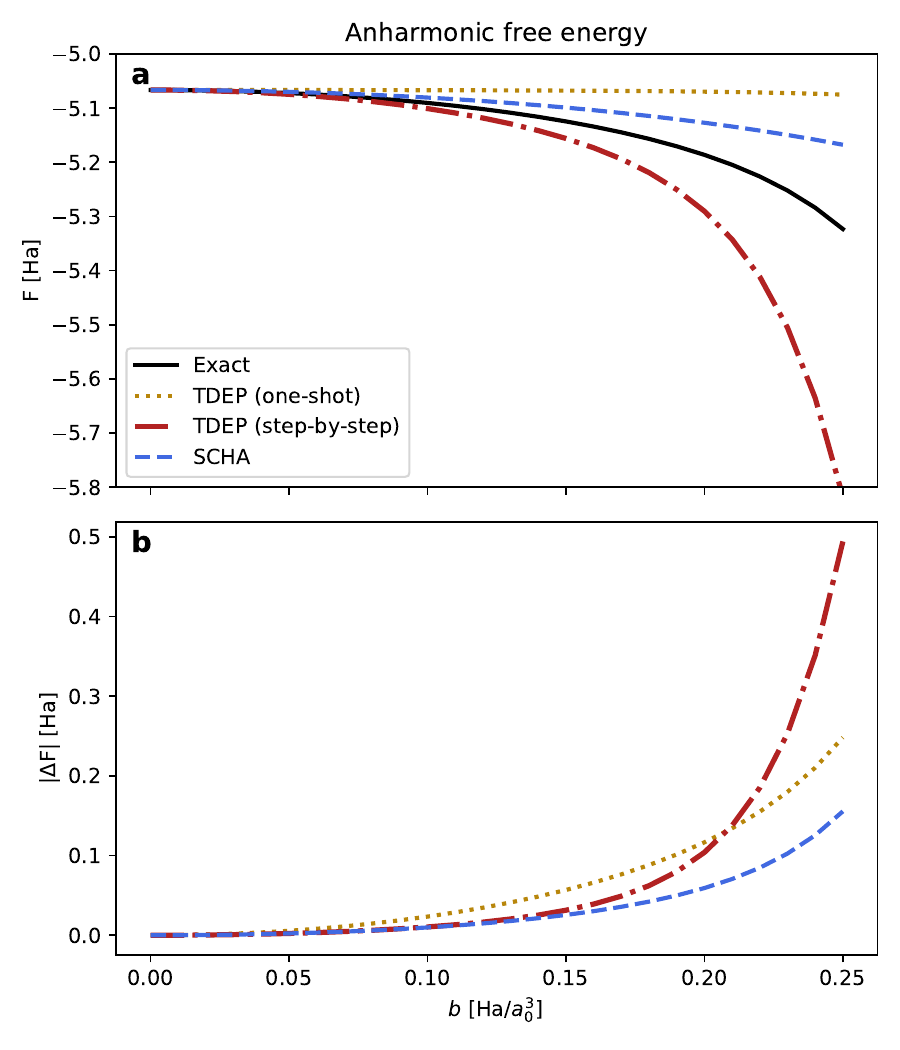}
    \caption{\small Free energy (\textbf{a}) of the potential \eqname~\eqref{eq:1dpot}, $\beta = \SI{1}{\hartree}$, comparing the exact (numerical) solution with SCHA and TDEP. Panel \textbf{(b)} reports the absolute value of the error versus the exact solution. The TDEP is implemented in two flavors: i) by fitting only the second-order force constant and then computing the free energy as in \eqname~\eqref{eq:tdep:start}, ii) by fitting up to the third-order force constant matrix and removing the high-order terms of the expansion to reduce the fluctuations, as reported in \eqname~\eqref{eq:v0:fluct} and \eqref{eq:free:tdep:wrong} (also 24-25 of ref.\cite{bottin_-tdep_2020}).}
    \label{fig:fe}
\end{figure}

The simulations are conducted for $\beta = \SI{1}{\hartree}$ ($E_\text{h}$ is Hartree atomic units) by varying the third-order anharmonicity $b$, and reported in \figurename~\ref{fig:fe}. Both the SCHA and TDEP \emph{step-by-step} offer variational expressions to the free energy following the inequalities presented in \secname~\ref{sec:scha:tdep} (\eqname~\ref{eq:free:scha} and \ref{eq:tdep:start}). When TDEP is employed to fit in one-shot also higher order force constants (third-order), it leads to much worse results, recovering the correct answer only in the harmonic potential ($b\to 0$). This may be surprising since the potential of \eqname~\eqref{eq:1dpot} is cubic, and the fit up to third-order force constants is exact. However, by inspecting the expression implemented a-TDEP, one implementation of the TDEP method\cite{bottin_-tdep_2020} for the free energy, the reason appears evident:
\begin{equation}
    \stackrel{(n)}{V} = \left<V(\bm r) - \sum_{p=1}^n \frac{1}{p!}\sum_{a_1\cdots a_p}\stackrel{(p)}{\Phi}_{a_1\cdots a_p}\prod_{k = 1}^p (R_{a_k} - \Rcalt_{a_k})\right>,
    \label{eq:v0:fluct}
\end{equation}
\begin{equation}
\stackrel{(n)}{F_\text{a-TDEP}} = \stackrel{(n)}{V} + \frac 32 Nk_BT - TS[\bPhit].
\label{eq:free:tdep:wrong}
\end{equation}
\eqname~\eqref{eq:free:tdep:wrong} does not tend to the exact free energy for $n\to\infty$ (it is not the cumulant expansion of the free energy). Therefore, there is no reason why increasing $n$ would improve the result's quality. Quite the opposite: in the $n\to\infty$ limit, TDEP (one-shot) fits the potential with the temperature-independent Taylor expansion. Then, \eqname~\eqref{eq:v0:fluct} coincides with the value of the potential energy in the average position
$$
\lim_{n\to\infty} \stackrel{(n)}{V} = V(\bRcalt),
$$
and the second-order fitted force constant $\bPhit$ coincides with the harmonic perturbative term of the Taylor expansion. Therefore, \eqname~\eqref{eq:free:tdep:wrong} becomes the free energy within the Harmonic approximation.
For the potential in \eqname~\eqref{eq:1dpot}, since $c$ is small, already a third-order fit is enough to recover the almost perfect harmonic limit, and the free energy obtained by this approach does not depend on the degree of anharmonicity $b$. The main reason for this failure is that the high-order terms fitted from the potential should be included in the expression for the entropy $S$ in \eqname~\eqref{eq:free:tdep:wrong}. A different TDEP implementation also accounted for the entropy perturbatively, likely restoring the perturbative expression of the free energy, but losing variationality (ref.\cite{benshalom_dielectric_2022}, SI IVb).
This work demonstrates that the TDEP method yields reliable results only when the fit is restricted to second-order force constants, as anticipated in \secname~\ref{sec:scha:tdep} (i.e. the step-by-step procedure). Notably, the common practice of fitting higher-order force constants in a one-shot remains widespread\cite{xie_first_principles2020,li_first-principles_2023,meitz_phonon_2025}, and it is even claimed in some cases to produce more accurate outcomes\cite{meitz_phonon_2025}. However, this approach lacks a formal foundation, as clearly shown here, and should be avoided, particularly for strongly anharmonic crystals.
The apparent improvement in heat capacity reported in ref.\cite{meitz_phonon_2025} is due to the use of an incorrect formula for the TDEP step-by-step heat capacity. Indeed, the heat capacity must be computed from the second temperature derivative of the variational TDEP free energy (\eqname~\ref{eq:var:tdep}), consistent with the procedure adopted in the SCHA framework\cite{monacelli_first-principles_2023}.
It is important to emphasize that the auxiliary force constant matrices obtained through the step-by-step TDEP fit are not a physical model of the interatomic energy landscape. Consequently, the energy of a configuration calculated using these auxiliary matrices is devoid of physical meaning. Using such energies to compute the variance or covariance along an MD trajectory leads to incorrect results for the heat capacity. The one-shot fit of the interatomic potential leads instead to a good approximation for energies of individual configurations, and thus is justified if employed to compute the heat capacity through the energy covariance across the MD trajectory\cite{meitz_phonon_2025}, explaining the origin of the better results obtained there for the one-shot procedure. In fact, this approach is very similar to fitting any empirical force field and evaluating the heat capacity directly from MD runs using the fitted interatomic potential.

Notably, although both the TDEP (step-by-step) and the SCHA satisfy a variational principle for the free energy, the SCHA exhibits superior performance in the presence of strong anharmonicity. This is likely due to the fact that the auxiliary SCHA phonons (squared eigenvalues of $\bPhis$) are generally harder than the TDEP auxiliary phonons (squared eigenvalues of $\bPhit$). This behavior arises from the connection between $\bPhit$ and the SCHA free energy Hessian (\secname~\ref{sec:static:response}), which, at leading order (\emph{bubble} approximation), always yields phonon frequencies that are softer than the auxiliary ones.
\begin{equation}
    F_\text{SCHA} = F_\bPhis + \left<V - \mathcal V_{\bRcals,\bPhis}\right>_{\bRcal,\bPhis},
\end{equation}
\begin{equation}
    F_\text{TDEP} = F_\bPhit + \left<V - \mathcal V_{\bRcalt,\bPhit}\right>_{\text{MD}}.
\end{equation}
Since the auxiliary phonons in $\bPhit$ are softer than those in $\bPhis$, the harmonic contribution to the free energy is lower in TDEP than in SCHA: $F_\bPhit < F_\bPhis$. As a result, the anharmonic correction term $\left<V - \mathcal V\right>$ must compensate more in TDEP than in SCHA, thereby amplifying the total error in systems with strong anharmonicity.

\begin{figure}
    \centering
    \includegraphics[width=0.5\columnwidth]{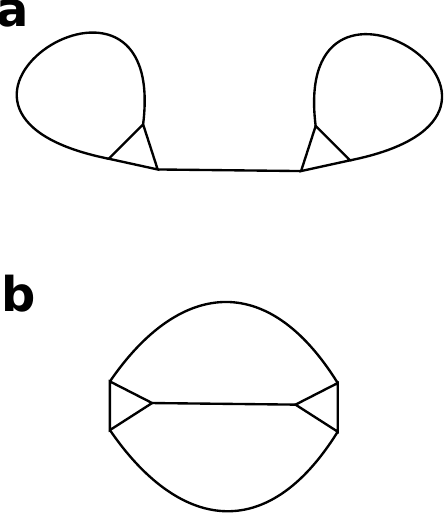}
    \caption{\small The two lowest-order Feynman diagrams for the free energy correction in the cubic anharmonicity. The triangles represent the three-phonon scattering vertex $\bPhithree$, while lines the one-phonon propagator. While \textbf{a} is accounted for within the SCHA, the \textbf{b} is not. The derivation and expression of these diagrams are reported in \appendixname~\ref{app:free:diagrams}.}
    \label{fig:fe:diagrams}
\end{figure}

\figurename~\ref{fig:fe} shows how both TDEP (step-by-step) and SCHA violate the perturbative limit of the free energy for $b\to 0$. In fact, the leading correction to the free energy is quadratic in the third-order anharmonicity $b$ and can be recast in the two Feynman diagrams reported in \figurename~\ref{fig:fe:diagrams} (derived in \appendixname~\ref{app:free:diagrams})

\begin{figure}
    \centering
    \includegraphics[width=\columnwidth]{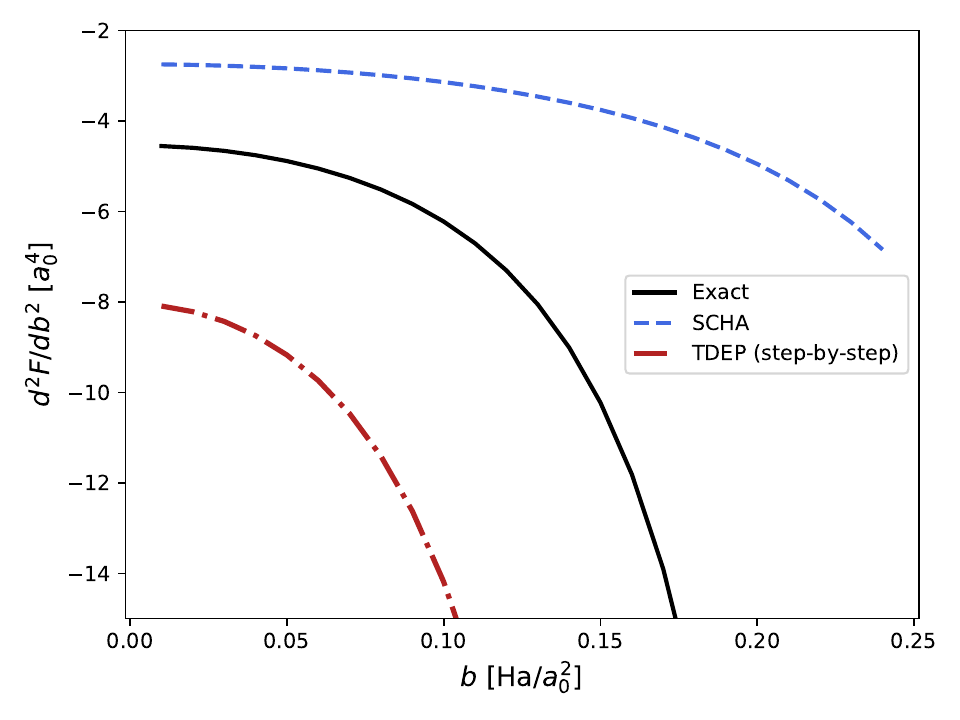}
    \caption{\small Curvature of the free energy with respect to the third order coefficient $b$. SCHA and TDEP fail to reproduce the correct free energy curvature on the anharmonicity in the $b\to 0$ limit.}
    \label{fig:fe:curv}
\end{figure}

As was discussed by ref.\cite{siciliano_wigner_2023}, the SCHA is limited to all possible Feynman diagrams in which each cut at time $t$ can intersect at most 2 phonon lines. In other words, the SCHA cannot describe more than 2 phonons propagating simultaneously throughout the system. While the one-phonon Green function does not have any of such diagrams at the lowest order of perturbation theory, making the response function correct in the perturbative limit, the same is not true for the free energy, where at the lowest order a diagram containing three simultaneous phonon lines contributes (\figurename~\ref{fig:fe:diagrams}\textbf{b}). This translates into a wrong curvature (second derivative for $b\to 0$) for the free energy as a function of anharmonic parameter $b$.
Indeed, the TDEP free energy also predicts the wrong curvature of the free energy, thus accounting for the two diagrams of \figurename~\ref{fig:fe:diagrams} with wrong prefactors.
The error on the curvature of both SCHA and TDEP, as shown \figurename~\ref{fig:fe:curv}, has an opposite sign for $b\to 0$, but it is not the same.
For this reason, computing the Free energy as the average between SCHA and TDEP, while in some cases it may improve the absolute value of the free energy, still fails in reproducing the perturbative limit and also comes at the cost of violating the variational expression, which is particularly useful in computing the phase diagram by assuring a (partial) error cancellation when comparing free energies of different phases.

\subsection{The static linear response theory - second-order phase transitions}

\label{sec:numerical:static}
This section compares the SCHA and various TDEP flavors to describe the static response function. Static response functions are fundamental to predict second-order phase transitions as the susceptibility diverges at the critical point. The static susceptibility of the model in \eqname~\eqref{eq:1dpot} is computed as described in  \secname~\ref{sec:static:response}: it is the derivative of the average atomic position upon the application of an external uniform force $\xi$ (\eqname~\ref{eq:response:function}).
The exact result has been evaluated numerically by computing the algorithmic differentiation of the average position $\left<R\right>_\xi$ by integrating with the trapezoid method over the distribution:
\begin{equation}
    p(R,\xi) = \frac{e^{-\beta \left[V(R) - \xi R\right]}}{\int_{-L}^L e^{-\beta \left[V(R) - \xi R\right]}}\qquad 
\end{equation}
where $L$ is set to \SI{10}{\bohr} to converge the result.

Perturbation theory on top of the harmonic result, when the fourth-order $c$ is small, leads to two corrections, quadratic in the third-order $b$, represented by the Feynman diagrams in \figurename~\ref{fig:chi:diagrams}\cite{bianco_second-order_2017}: the \emph{tadpole} (\figurename~\ref{fig:chi:diagrams}\textbf{a}) and the \emph{bubble} (\figurename~\ref{fig:chi:diagrams}\textbf{b}).

\begin{figure}
    \centering
    \includegraphics[width=\columnwidth]{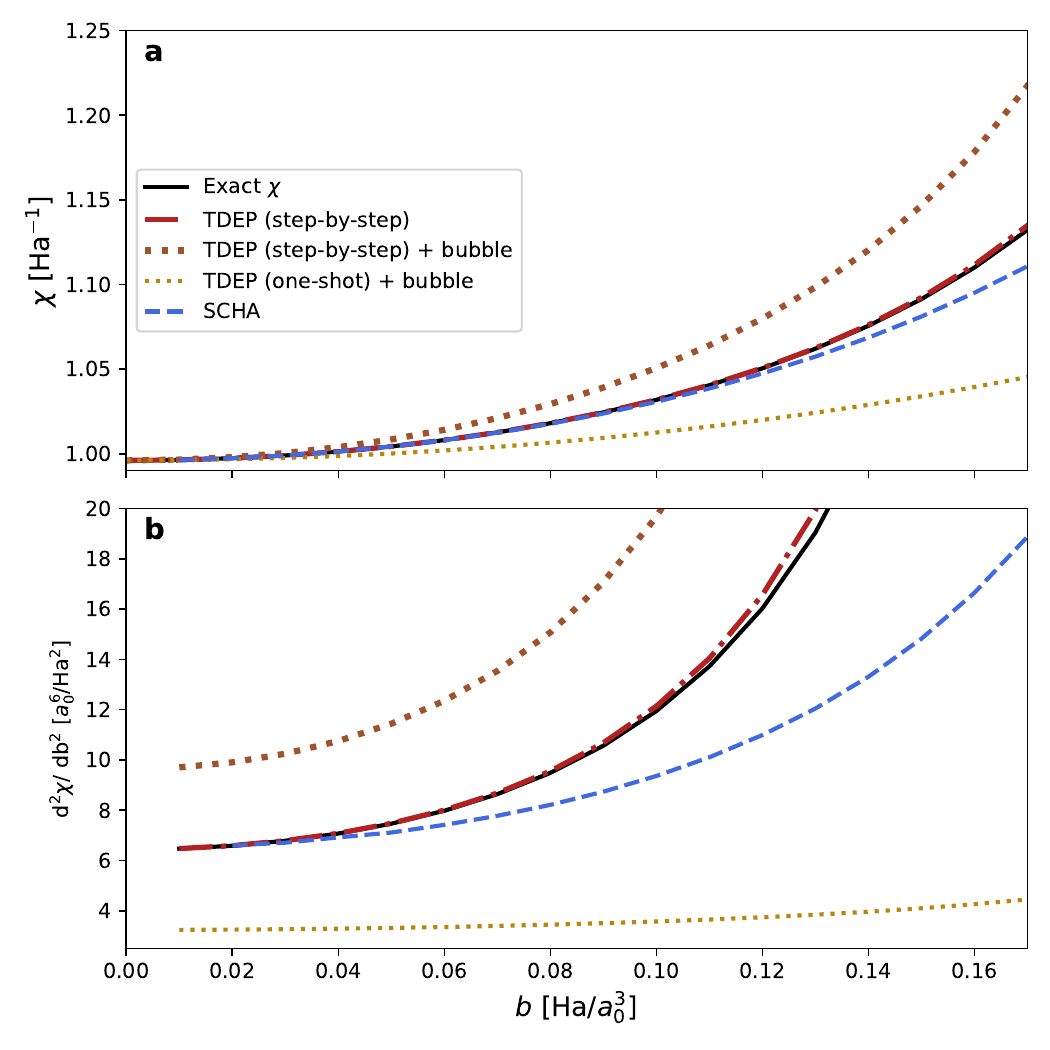}
    \caption{\small Comparison of the static response function between TDEP, SCHA and the exact result. Panel \textbf{a} reports the susceptibility $\chi$ as a function of the anharmonicity. Panel \textbf{b} shows the second derivative of $\chi$ for the third-order anharmonic parameter $b$. Only SCHA and TDEP (second-order) present the correct $b\to 0$ limit, demonstrating how they are the only two approaches that correctly reproduce the perturbative result.}
    \label{fig:static:response}
\end{figure}

\figurename~\ref{fig:static:response} compares the static susceptibility from SCHA, TDEP (step-by-step), and TDEP (one-shot) with the exact result.
As discussed in \secname~\ref{sec:static:response}, TDEP (step-by-step) auxiliary force constant matrix $\bPhit$ matches the exact result independently of the degree of anharmonicity $b$. This is a consequence of the fluctuation-dissipation theorem, as $\bPhit$ coincides with the inverse of the covariance matrix. \figurename~\ref{fig:static:response}\textbf{b} shows the perturbative limit of the response function. 
The SCHA and the TDEP (step-by-step) static susceptibilities have the correct curvature versus the $b$ parameter (perturbative limit). This was already demonstrated for the SCHA in ref.\cite{bianco_second-order_2017} and, to my knowledge, never proved for TDEP (step-by-step). 

\begin{figure}
    \centering
    \includegraphics[width=\columnwidth]{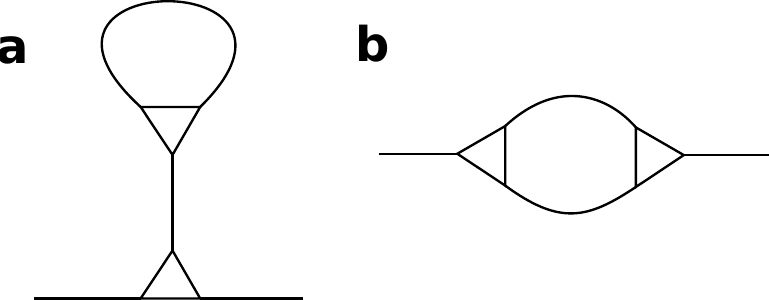}
    \caption{\small Lowest order diagrammatic correction to the static response function. The triangle represents the three-phonon scattering vertex $\bPhithree$, while a solid line is the phonon propagator. Panel \textbf{a} reports the \emph{tadpole} diagram, responsible for the change of the centroids $\left<r\right>$ due to anharmonicity (real self-energy). Panel \textbf{b} displays the \emph{bubble} diagram, whose dynamical equation is reported in \eqname~\ref{eq:diag:bubble:three}. Both diagrams are quadratic in $\bPhithree$, and the lowest order correction of anharmonicity to the static and dynamic susceptibilities.}
    \label{fig:chi:diagrams}
\end{figure}
The same \figurename~\ref{fig:static:response}\textbf{b} also shows how only the step-by-step approach to TDEP does not fail.
As discussed in \secname~\ref{sec:static:response}, the TDEP force constant matrix $\bPhit$ is always positive definite, as it is equal to the position-position static correlator, even if the centroid positions are constrained to an unstable position. Therefore, exploiting $\bPhit$ to assess a structure's stability is impossible. For this reason, a bubble diagram contribution is sometimes added to $\bPhit$  (\figurename~\ref{fig:chi:diagrams}\textbf{b}) in the hope of describing a transition to a saddle point of the free energy landscape to disentangle the stability of a structure.
For example, in ref.\cite{dangic_origin_2021} the TDEP $\bPhit$ is corrected with the static self-energy of the \emph{bubble} diagram $\boldsymbol{\Pi}(\omega\to 0)$ (\eqname~\ref{eq:diag:bubble:three}) to assess the phase transition between rhombohedral and cubic \ch{GeTe}.
However, this procedure leads to wrong results as proved in \secname~\ref{sec:static:response}: adding the extra bubble diagram leads to overcounting anharmonicity. The absence of a divergence in the bare TDEP susceptibility $\bPhit$ observed in ref.\cite{dangic_origin_2021} is either a sign of a nondisplacive phase-transition in \ch{GeTe} or a manifestation of the difficulties in converging TDEP results in the proximity of a phase-transition, with the latter being more likely as confirmed by a subsequent analysis from some of the same authors only relying on more expensive but exact MD correlation functions\cite{dangic_2022_molecular}.

As we discussed for the free energy, the TDEP (one-shot) procedure leads to wrong results.  
In this case, the anharmonicity is undersampled, as the resulting second-order force constant matrix matches with the harmonic force-constant matrix (due to the negligible value of $c$ in \eqname~\ref{eq:1dpot}). The only correction to evaluate the susceptibilities from the harmonic susceptibilities by TDEP comes from the \emph{bubble} diagram (\figurename~\ref{fig:chi:diagrams}\textbf{b}), thus missing the role played by the \emph{tadpole} (\figurename~\ref{fig:chi:diagrams}\textbf{a}) and underestimating the perturbative limit (\figurename~\ref{fig:static:response}\textbf{b}).

\subsection{Dynamical spectral functions}
\label{sec:test:dynamical}
One of the most promising applications of SCHA and TDEP is the calculation of dynamical spectral functions, which play a dominant role in estimating the lattice thermal conductivity.

The exact dynamic response function is evaluated by numerically solving the Liouville equation (\eqname~\ref{eq:liouville}). The equilibrium solution is perturbed with a time-dependent potential $V'(x, t) = x\delta(t - t_0)\Delta$, where $\delta(t-t_0)$ is the Dirac's delta function, here employed to excite all frequencies with the same amplitude.
This results in a shift of the momentum $p$ of the equilibrium distribution by a (small) finite value $\Delta$
\begin{equation}
    \rho(x, p, t_0^+) = \rho(x, p + \Delta, t_0^-) 
\end{equation}
\begin{equation}
    \rho(x, p, t_0^+) =\frac{\exp\left[-\beta\frac{\left(p + \Delta\right)^2}{2m} - \beta V(x )\right]}{Z}.
    \label{eq:initial:condition}
\end{equation}
Then, the response function is evaluated as the Fourier transform of the average position as a function of time (\eqname~\ref{eq:avgpos} and \ref{eq:green}).
The Liuville equation is solved with finite differences on a grid between \SI{-4}{\bohr} and \SI{4}{\bohr} discretized by $\Delta x = \SI{0.05}{\bohr}$, $\Delta p = \SI{0.05}{\per\bohr}$, and $\Delta t = \SI{2.5e-3}{\per\hartree}$, with a perturbation $\Delta = \SI{0.1}{\bohr^{-1}}$.  Here, we fixed $b=\SI{0.21}{\hartree\per\bohr^3}$.

\begin{figure}
    \centering
    \includegraphics[width=\columnwidth]{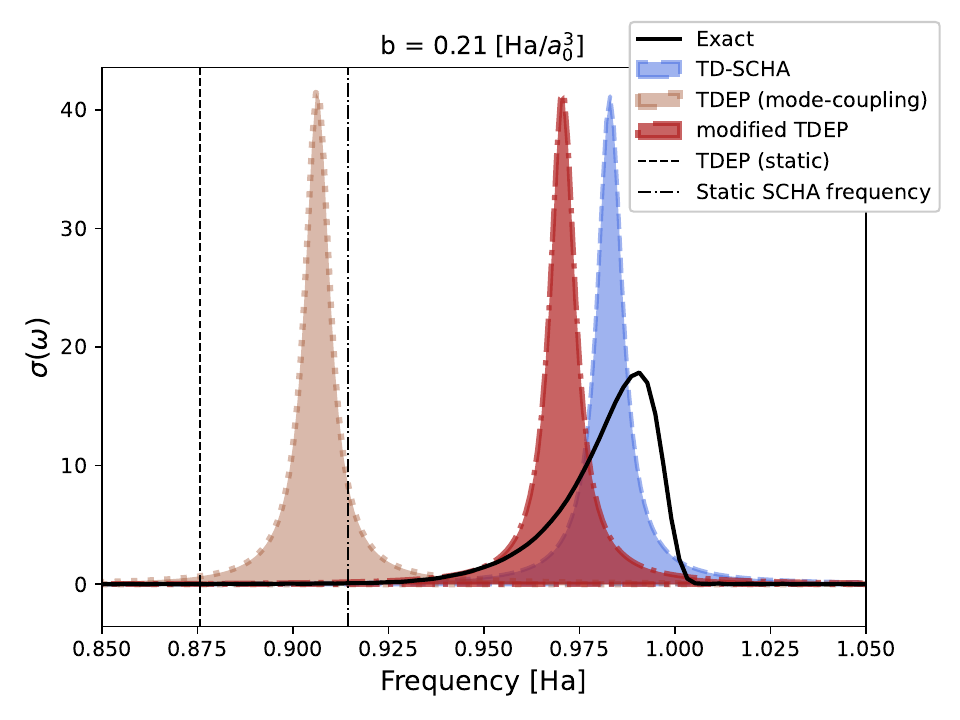}
    \caption{\small Dynamical spectral function of the 1D potential \eqname~\eqref{eq:1dpot} where $b=\SI{0.21}{\hartree\per\bohr^3}$ and $k_BT=\SI{1}{\hartree}$. All the spectral functions with TDEP and TD-SCHA have an imaginary smearing of $\eta = \SI{4e-3}{\hartree}$. 
    The vertical dashed (dash-dot) lines represent the pole of the static TDEP (SCHA) response function.}
    \label{fig:dynamical:spectral}
\end{figure}

\figurename~\ref{fig:dynamical:spectral} reports the comparison between the exact spectral function and the one obtained by TDEP (step-by-step + bubble, as introduced in the mode-coupling theory\cite{castellano_mode-coupling_2023}, \eqname~\ref{eq:chit}), TD-SCHA (\eqname~\ref{eq:chis}), and the modified TDEP introduced in this work to correct the static shift of the self-energy introduced by the wrong boundary conditions on the Krames-Kronig relations (\eqname~\ref{eq:dyn:tdep:fixed}), see \secname~\ref{sec:dynamical} for more details.

From \figurename~\ref{fig:dynamical:spectral}, the TDEP (mode-coupling)\cite{castellano_mode-coupling_2023} shows a global red-shift of the frequencies, leading to a worse agreement than TD-SCHA with the exact result. This red shift is compensated when the static \emph{bubble} contribution is removed from the self-energy (modified TDEP, \eqname~\eqref{eq:dyn:tdep:fixed}), and a much closer agreement is obtained, demonstrating the effectiveness of the correction introduced in this work. 
Another interesting feature depicted in \figurename~\ref{fig:dynamical:spectral} is the distance between the TD-SCHA and the TDEP spectral function peak and their respective static susceptibility pole. This difference originates from the real part of the \emph{bubble} self-energy that depends on frequency, and it changes sign at finite frequencies compared to $\omega\to 0$. Lastly, the spectrum's shape for all the approximate theories does not deviate from a Lorentzian, as displayed by the exact result. The asymmetric peak shape is due to the continuous population of states in an anharmonic oscillator, whose energy difference decreases with increasing quantum number. Both the \emph{bubble} approximation and the full RPA of TD-SCHA cannot recover this behaviour. Indeed, in a solid with multiple phonons, both approximations can produce a non-Lorentzian linewidth due to the continuum of possible phonon-phonon scattering, which is not present in a molecule (see \secname~\ref{sec:real:systems}).

\begin{figure}
    \centering
    \includegraphics[width=\columnwidth]{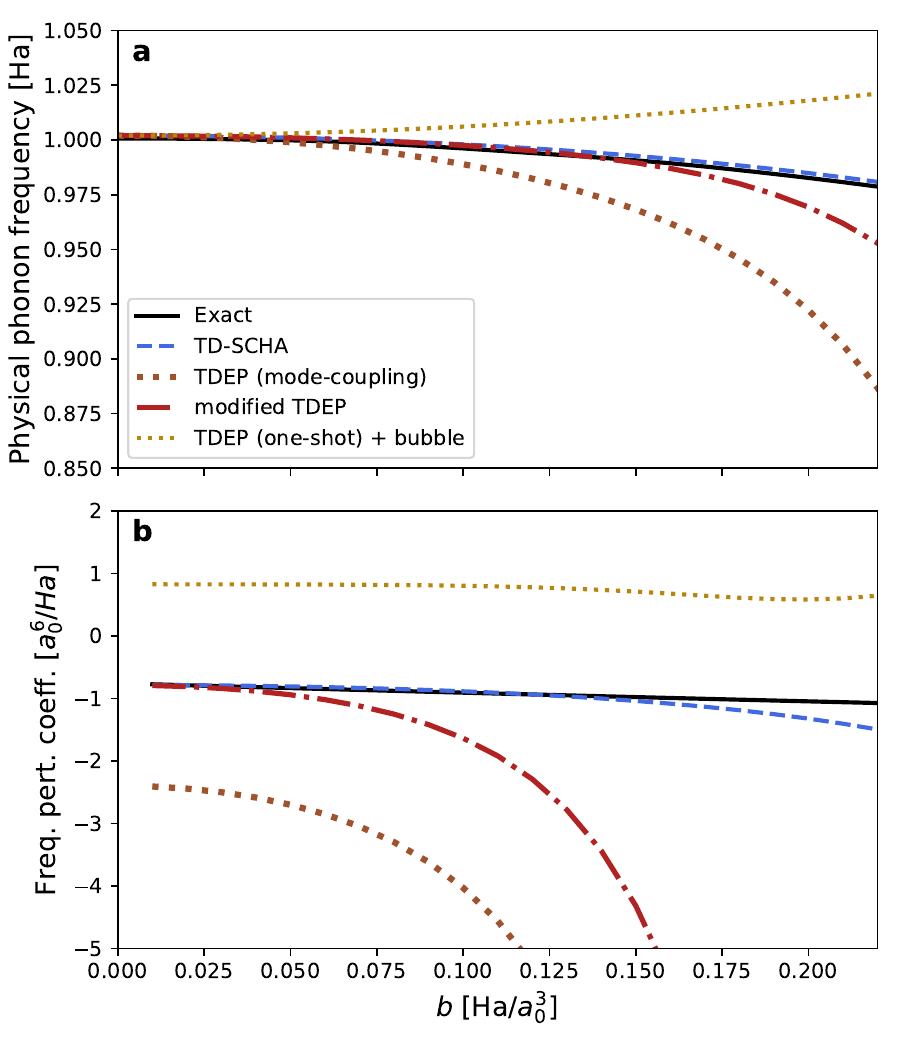}
    \caption{\small Dynamical frequency computed as the pole of the spectral function $\sigma(\omega)$ (\figurename~\ref{fig:dynamical:spectral}) comparing different approximations (potential reported in \eqname~\ref{eq:1dpot}) Panel \textbf{a}: Poles of the spectral functions. Panel \textbf{b}: second-derivative of the poles as a function of the anharmonicity $b$. Lines matching the poles for $b\to 0$ indicate a correct perturbative limit.}
    \label{fig:dynamical:poles}
\end{figure}

To further investigate the discrepancies with the exact results of the various approximations, \figurename~\ref{fig:dynamical:poles}\textbf{a} compares the average position of the spectral peak (the pole of the spectral function) as a function of the third-order anharmonicity (the $b$ parameter). 
As always, all the methods give the exact result in the harmonic limit  ($b\to 0$). However, as we introduce anharmonicity, only the TD-SCHA and the modified TDEP here introduced provide the correct perturbative coefficient. This is well shown by \figurename~\ref{fig:dynamical:poles}\textbf{b}, where the second derivative of the dynamical average frequency with $b$ is reported.
The TDEP mode-coupling approach\cite{castellano_mode-coupling_2023} (step-by-step, adding the dynamical bubble self-energy) violates the perturbative limit by double-counting the static bubble self-energy, which is already accounted for by $\bPhit$. The agreement of TDEP is, in fact, much improved if the static limit ($\omega\to 0$) of the bubble contribution is removed from the self-energy, restoring the correct boundary condition $\bPit(0) = 0$ for the Kramers-Kronig relations, as done in the modified TDEP (\eqname~\ref{eq:dyn:tdep:fixed}).
This result contrasts with the claim of ref.\cite{castellano_mode-coupling_2023} that the mode-coupling TDEP satisfies the perturbative limit, which is untrue, unless the modified TDEP here introduced is accounted for.
Notably, the TDEP (one-shot) dressed with the \emph{bubble} self-energy performs even worse than the harmonic approximation, missing even qualitatively the sign of the peak shift introduced by the anharmonicity. 

Despite this critical discrepancy in the position of the dynamical vibrational frequency, all of these methods still give the correct perturbative value for the imaginary part of the self-energy, thus predicting reasonable values for the phonon lifetimes.
In fact, phonon lifetimes depend on the imaginary part of the \emph{bubble} self-energy, which is nonzero only at finite frequency. This means that all the approaches discussed so far correctly evaluate the imaginary part of the self-energy in the perturbative scheme, as it is equivalent to the Fermi Golden rule.

This is probably the reason for the success of all different implementations of TDEP in computing thermal transport properties, where most of the contribution is given by the phonon lifetimes related to the imaginary part of the self-energy.
However, in strongly anharmonic systems outside the perturbative regime, the real part of the self-energy also plays a fundamental role, especially in the proximity of critical points. In this case, differences in the phonon frequencies arising from the real part of the self-energy may hamper the accurate estimation of thermal transport properties, and the correction proposed in \eqname~\eqref{eq:dyn:tdep:fixed} becomes essential.

\begin{figure*}
    \centering
    \includegraphics[width=0.95\textwidth]{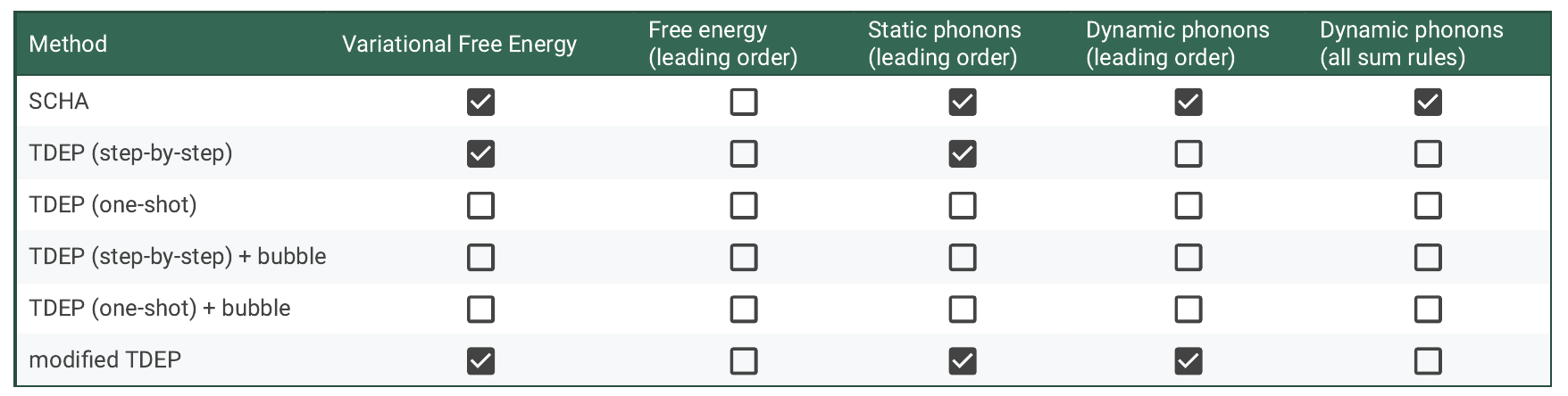}
    \caption{Comparison between SCHA and several TDEP implementations: step-by-step consists in fitting high-order anharmonic tensors on the residual forces of the quadratic fit (\eqname~\ref{eq:tdep:fit:3}), one-shot in fitting a high-order anharmonic tensor together with the quadratic one. ``+ bubble'' means that the bubble self-energy is added to the TDEP dynamical matrix $\bPhit$. The modified TDEP is the approach introduced in this work (\eqname~\ref{eq:dyn:tdep:fixed}). The columns address whether the expression of the free energy is variational, if the free energy, static, and dynamic susceptibilities are correct to the leading order in the anharmonicity, and whether the dynamical expression of the self-energy satisfies all the sum rules (only true in the full RPA of TD-SCHA).
    None of the theories addressed here provides free energy that is correct in the leading order of anharmonicity. However, it is possible to manually add the missing term (\figurename~\ref{fig:fe:diagrams}) to the free energy expression in both SCHA and TDEP. However, this introduction spoils the variationality.
    }
    \label{fig:table}
\end{figure*}

In \figurename~\ref{fig:table}, we summarize all the features of SCHA and different TDEP implementations.

\section{Self-consistent Phonons}
\label{sec:scp}

The Self-Consistent Phonons (SCP) approach originates from the many-body theory and has a very long history\cite{Born1912}; it is the phononic equivalent of Hartree-Fock for electrons.
It solves the Dyson equation for the phonons' Green function self-consistently by truncating the self-energy to a given order.
The method comes in multiple flavors depending on the processes accounted for in the self-energy.

\begin{figure}
    \centering
    \includegraphics[width=0.7\columnwidth]{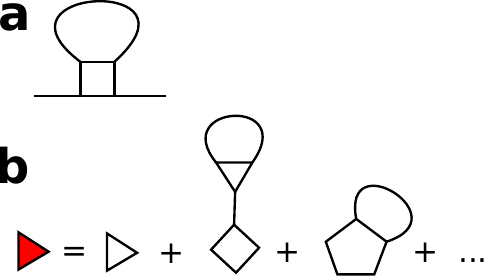}
    \caption{\small Anharmonic diagrams that can be accounted for in the SCP theory. Panel \textbf{a} reports the \emph{loop} diagram: the lowest order correction of the four-phonon scattering vertex $\bPhifour$. Panel \textbf{b} shows how phonon-phonon scattering vertices are re-normalized, accounting for higher-order terms in the SCHA. Tadpole-like corrections appended to one vertex account for the change of average position due to third-order anharmonicity, while a loop connecting two vertices accounts for the effect that higher-order terms of the phonon-phonon interaction have when taking the average (\eqname~\ref{eq:phi3} and \ref{eq:phi4})}
    \label{fig:diags:scp}
\end{figure}

The simplest and most common implementation of SCP consists in truncating the self-energy only to the \emph{loop} diagram (\figurename~\ref{fig:diags:scp}\textbf{a}), and assuming that it commutes with the harmonic dynamical matrix (i.e., it is diagonal in the basis of the harmonic phonon polarization vectors). This assumption is called the \emph{mode-mixing}. Notably, while in principle it is possible to use this approximation both within the SCHA and TDEP, the implementations here discussed go beyond the \emph{mode-mixing}, allowing the complete relaxation of the $\bPhis$ and $\bPhit$ force-constant matrices, including their eigenvectors. The \emph{mode-mixing} introduced for TDEP (and sometimes SCHA) is different as it only regards the dynamical self-energy (see \secname~\ref{sec:dynamical}), and is applied for diagrams that have a nonvanishing imaginary part, like the \emph{bubble}.

The self-consistency originates from the propagator that loops on the four-phonons scattering operator $\bPhifour$, which depends on the $t\to 0^+$ one-phonon Green function:
\begin{equation}
    \Phi_{ab}^\text{\tiny(SCP)} = \Phi^{(0)}_{ab} + \Pi_{ab}(\bPhi^\text{\tiny(SCP)})
\end{equation}
\begin{equation}
    \Pi_{ab}(\bPhi^\text{\tiny(SCP)}) = \frac 12\sum_{cd} \Phifour_{abcd}\left<(R_c - \Rcal_c)(R_d - \Rcal_d)\right>_{\bPhi^\text{\tiny(SCP)}}.
\end{equation}
SCP was successfully applied in solid helium\cite{Klein1972}, where, due to the high symmetry of the structure, only the frequencies of the $\bPhi^\text{\tiny(SCP)}$ must be updated in the self-consistent cycle if the mesh in the Brillouin zone only includes high-symmetry points.
Interestingly, this approach reduces itself to the SCHA in a crystal where all the lattice positions are constrained by symmetry (eliminating the \emph{tadpole} diagram in \figurename~\ref{fig:chi:diagrams}\textbf{a}), and if the interatomic potential is at most quartic. However, even in this limit, the phonons determined by SCP considering only the loop still violate the perturbative limit as no symmetry argument can cancel out the bubble diagram (\figurename~\ref{fig:chi:diagrams}\textbf{b}). For similarity with perturbation theory, the bubble diagram is often added \emph{a posteriori} after the self-consistency of the loop is achieved.
This has been employed for the first time to study the phonon spectra of solid neon\cite{Koehler1969}, and more recently used, e.g., in the simulation of \ch{SrTiO3} cubic phase\cite{tadano_self-consistent_2015}, where all the parameters of the lattice are constrained by the $Pm\bar 3m$ symmetry group, thus leading to a result that correctly reproduces the perturbative limit both in $\bPhifour$ and in $\bPhithree$. Indeed, the SCP method can be easily improved to add the contribution of diagram \figurename~\ref{fig:fe:diagrams}\textbf{b} also a posteriori for the calculation of the free energy\cite{Oba_2019}, and to relax the hypothesis of crystal lattice position fully constrained by symmetry\cite{Masuki_2022} by adding the \emph{tadpole} in the expression of the self-consistent self-energy (\figurename~\ref{fig:chi:diagrams}\textbf{a}) and by renormalizing the anharmonic scattering vertices based on the relaxed structure (\figurename~\ref{fig:diags:scp}\textbf{b}) by dressing high-order scattering vertices with tadpole diagrams. 

Given its diagrammatic origin, this method is built on a rigorous perturbative expansion of the solution with anharmonicity. Therefore, it leads to a result that matches perturbation theory up to any arbitrary order, thus beating the limits of TDEP and SCHA encountered in this work.
The other side of the coin is that correcting high-order terms comes at an exponentially high computational cost, thus limiting its practical applicability to only describing the leading order corrections. 

An intrinsic limitation of the SCP method is that it remains bound to the Taylor expansion of the interatomic potential, which becomes significantly more expensive as the order increases. Moreover, the SCP Green's function and susceptibilities do not emerge from a variational approach (except when the self-energy is truncated to reproduce the SCHA, for which the two methods coincide).
This has two consequences that must be considered carefully when employing this methodology: i) nonvariational free energy does not necessarily lead to error cancellation across different phases, and increasing the self-energy's order does not necessarily convert into a more accurate phase diagram. ii) A truncation in the diagrammatic expansion of the self-energy may violate some known symmetry properties of materials. A noteworthy example arises from the rotational sum-rule violation of the first-order SCP theory, which occurs in freestanding 2D materials, where the truncation of the self-energy leads to a linear dispersion of the flexural mode around the center of the Brillouin zone. In contrast, by symmetry arguments, the exact solution must exhibit a quadratic dispersion\cite{Aseginolaza2024}, which is recovered by the SCHA when considering the complete RPA expansion. 

\section{Benchmarks on PbTe and CsSnI3}
\label{sec:real:systems}
In this section, we benchmark TDEP and SCHA on two prototypical materials known for their strongly anharmonic properties: lead telluride (PbTe) and the metal halide perovskite \ch{CsSnI3}. Both materials have been investigated using both techniques. 

\ch{PbTe} is a paraelectric material (does not develop a bulk polarization) down to \SI{0}{\kelvin}, often regarded as an incipient ferroelectric\cite{delaire_giant_2011}, as even a small strain is sufficient to induce a ferroelectric transition\cite{jiang_origin_2016}.
It crystallizes in a rocksalt high-symmetry structure. Its anharmonic properties were first studied using the TDEP method\cite{romero_thermal_2015}, which unveiled the peculiar phonon spectrum around $\Gamma$ that shows a satellite peak correctly identified in the experiments\cite{delaire_giant_2011}. Later, SCHA calculation reported an excellent agreement with the TDEP and the experimental realization\cite{ribeiro_strong_2018}. Here, we investigate the origin of this agreement found in \ch{PbTe} despite the differences discussed in both methods, and when we expect these differences to emerge.

\begin{figure*}
    \centering
    \includegraphics[width=0.9\textwidth]{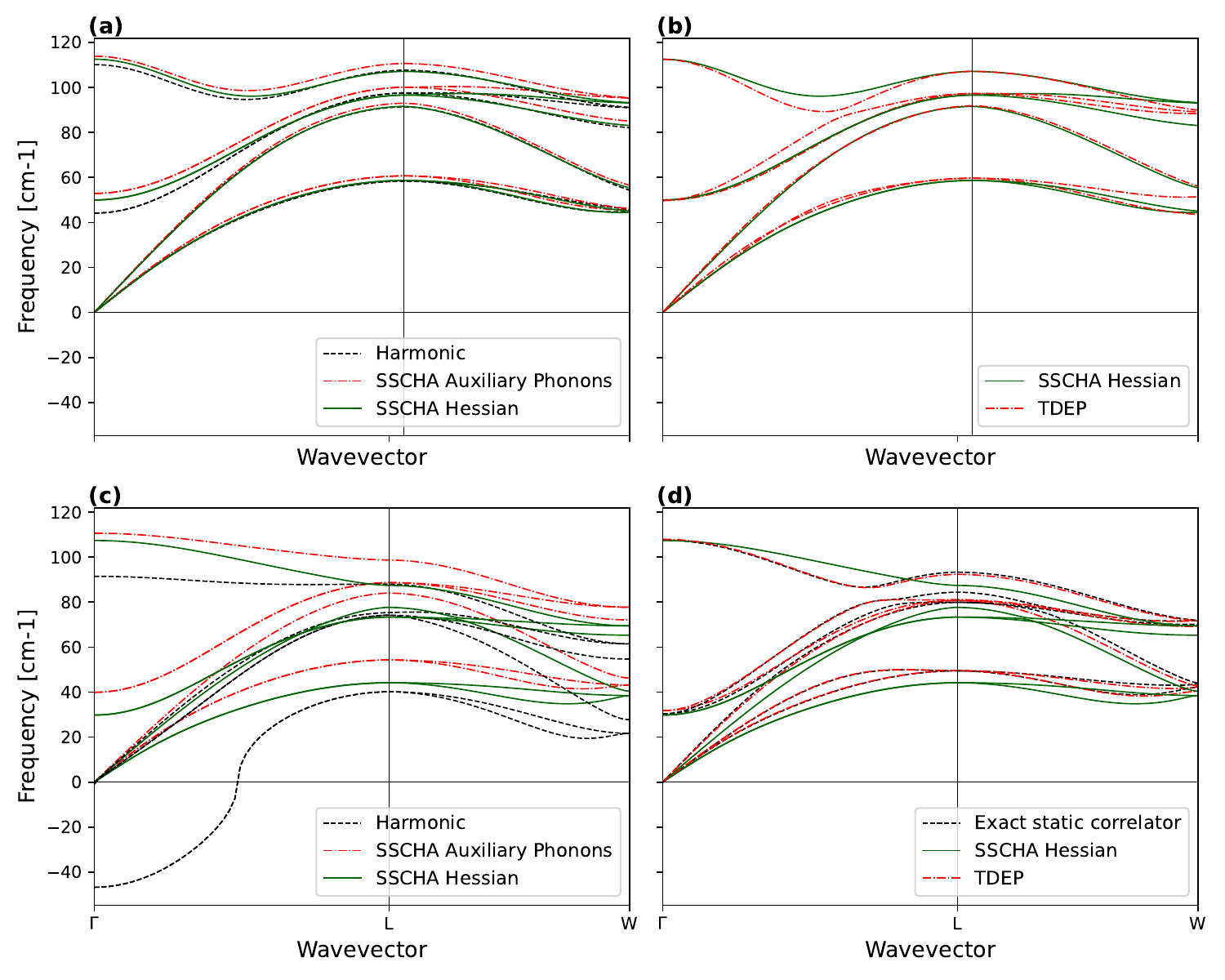}
    \caption{\ch{PbTe} static phonon spectrum simulated within SCHA and TDEP at \SI{300}{\kelvin} to suppress quantum effects. \textbf{(a)} and \textbf{(b)} are for \ch{PbTe} at the equilibrium volume (LDA), as performed in ref.\cite{ribeiro_strong_2018}. \textbf{(c)} and \textbf{(d)} report the phonon dispersion when the volume has been expanded by \SI{-3.1}{\giga\pascal} to enhance the effect of anharmonicity. Despite the anharmonicity unveiled by the dynamical \emph{bubble} self-energy, at the LDA volume, Harmonic, SCHA auxiliary, and SCHA hessian phonons are very similar, explaining the consistency with the TDEP result. More significant differences are observed in the strained sample, where the auxiliary phonons obtained by TDEP and SCHA deviate (\textbf{c}), while the free energy SCHA Hessian is in good agreement with the TDEP and the exact static phonon dispersions (\textbf{d}).}
    \label{fig:pbte}
\end{figure*}

\figurename~\ref{fig:pbte} shows the phonon dispersion at \SI{300}{\kelvin} for \ch{PbTe} at ambient conditions and in a strained cell (\SI{-3.1}{\giga\pascal}) that enhances the anharmonicity (where Harmonic phonons are imaginary). In the first case, it is evident that, despite the strong anharmonicity giving rise to a satellite peak reported in the spectral function\cite{romero_thermal_2015,ribeiro_strong_2018}, the real part of the bubble self-energy does not alter the phonon frequencies, as the SCHA auxiliary and Hessian phonons are very similar. This explains the similarity of the result obtained between the SCHA and TDEP, as the shift provided by the real part of the bubble self-energy is surprisingly negligible in this system, even in the presence of strong anharmonicity. For this reason, the modified TDEP approach leads to results in agreement with the mode-coupling TDEP.

Differences emerge when the lattice is strained, and the Harmonic dispersion becomes imaginary. In this state, the Hessian differs from SCHA auxiliary modes, pinpointing a non-negligible role played by the static bubble self-energy.
Therefore, properly accounting for the correction of self-energy on top of TDEP and SCHA is fundamental here.

Another key difference between the TDEP and SCHA spectral functions lies in the treatment of the dynamical self-energy. Specifically, TDEP approximates the self-energy by retaining only the \emph{bubble} diagram\cite{castellano_mode-coupling_2023}, whereas the TD-SCHA, in principle, includes the entire series of RPA diagrams\cite{lihm2020gaussian,siciliano_wigner_2023} (\figurename~\ref{fig:self:energy}). 
Nevertheless, due to the computational challenges associated with interpolating the full RPA series on a dense $\boldsymbol{q}$-mesh in the Brillouin zone, the dynamical self-energy is often truncated to the \emph{bubble} term even within the SCHA framework\cite{monacelli_stochastic_2021}. In this work, we assess the validity of this approximation, which has not been systematically investigated to date. 
While the \emph{bubble} approximation is typically adequate for most condensed matter systems, recent findings have identified materials where it breaks down. One such example is \ch{CsSnI3}, a prototypical metal-halide perovskite that has attracted significant attention for its potential in photovoltaic and thermoelectric applications. At ambient pressure, \ch{CsSnI3} exhibits a complex phase diagram, transitioning from an orthorhombic ($\gamma$) to tetragonal ($\beta$), and finally to a cubic ($\alpha$) phase at \SI{360}{\kelvin} and \SI{440}{\kelvin}, respectively\cite{Chung2012}.

\begin{figure*}
    \centering
    \includegraphics[width=\textwidth]{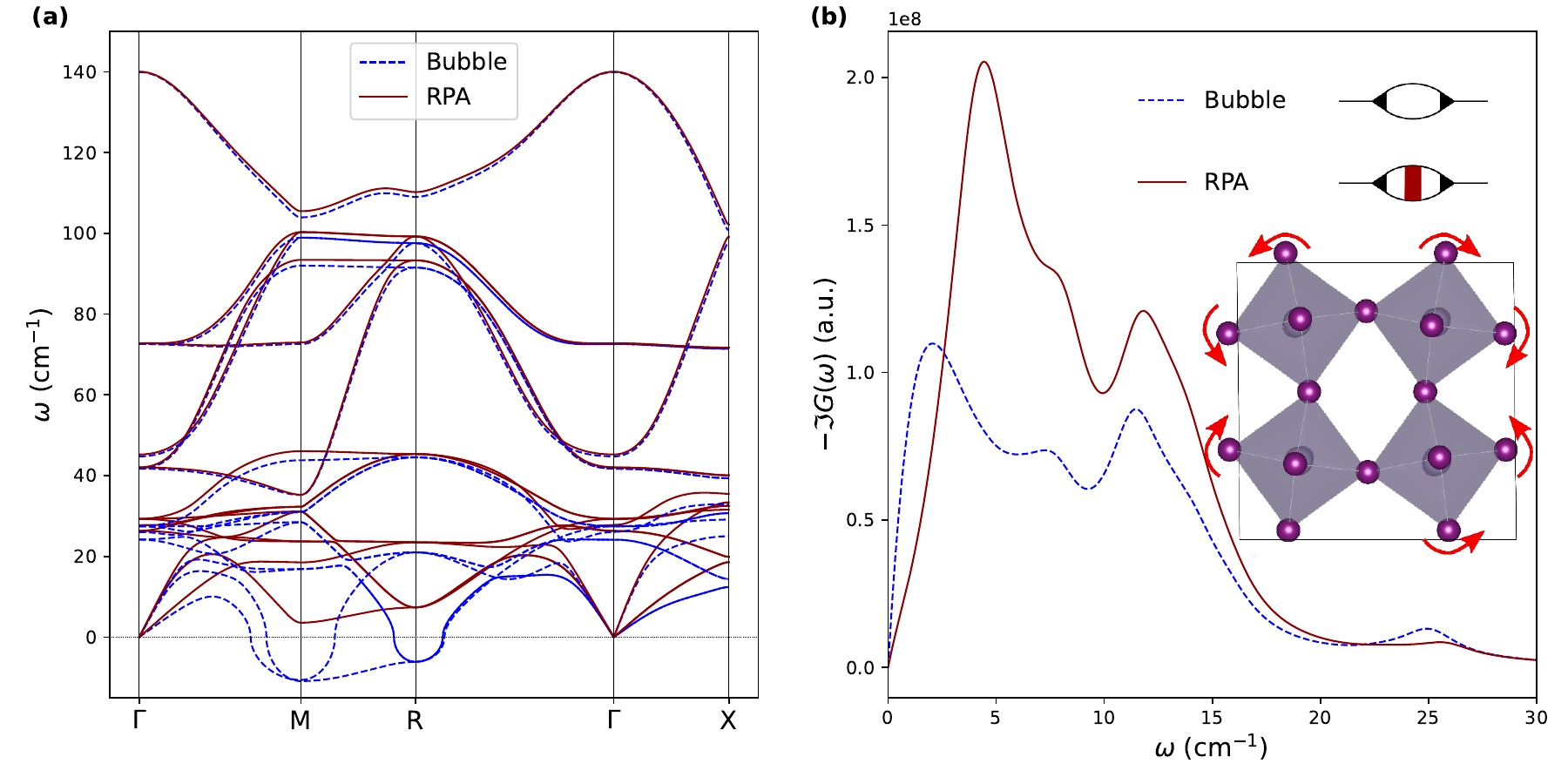}
    \caption{Static and dynamical spectral functions of \ch{CsSnI3} at \SI{450}{\kelvin}. \textbf{(a)} Poles of the static spectral function, i.e., the anharmonic phonon frequencies from the Hessian of the free energy, computed within the SCHA at the \emph{bubble} approximation and the full RPA. \textbf{(b)} Dynamical spectral function of the soft phonon mode at M. The inset displays the atomic motion associated with this mode. The bubble approximation overestimates anharmonicity, leading to a fake instability from the appearance of imaginary frequencies. The higher-order diagrams accounted for by the RPA cancel out part of the anharmonic renormalization from the bubble, leading to a stable cubic structure at \SI{450}{\kelvin}. This is reflected in the dynamic spectral function, where the bubble approximation misses the main resonance of the spectrum.}
    \label{fig:CsSnI3}
\end{figure*}

In \figurename~\ref{fig:CsSnI3}, we show the result of the free energy Hessian (panel \textbf{a}) and the spectral function (panel \textbf{b}) within the TD-SCHA at \SI{450}{\kelvin} for the cubic ($\alpha$) phase, where the interatomic potential of \ch{CsSnI3} is computed from first principles via density functional theory and the PBEsol exchange correlation potential (see ref.\cite{monacelli_first-principles_2023} for more details on the DFT calculations). 
Here, the \emph{bubble} approximation for the free energy Hessian agrees with the full RPA for all phonon bands, except for the lowest energy one, which is imaginary within the bubble (unstable) and real within the RPA (stable). Indeed, \ch{CsSnI3} is cubic at \SI{450}{\kelvin}, and the tetragonal phase spontaneously relaxes into the cubic cell, as shown in ref.\cite{monacelli_first-principles_2023}. The TDEP force constant matrix is expected to reproduce the behaviour of RPA and better, as it is exact (except for quantum effects, which are negligible at \SI{450}{\kelvin} in this system). 

A similar behavior is observed in other isostructural metal-halide perovskites, such as \ch{CsPbBr3}\cite{Fransson2023}, where TDEP (referred to as EHM in that work) accurately reproduces the static frequencies of the M-tilt mode. In contrast, the auxiliary phonon frequencies obtained from SCHA (and similarly from SCP) are significantly overestimated.

Including the \emph{bubble} diagram within the SCHA and SCP frameworks leads to an incorrect unstable phonon mode, as shown in \figurename~\ref{fig:CsSnI3}\textbf{(a)}. Only the full RPA inversion yields the correct static correlation function. Interestingly, the same authors report a better agreement between the MD-derived dynamical correlation function and the auxiliary phonon spectrum from SCHA (or SCP) for the highest optical phonon branch, compared to TDEP (ref.\cite{Fransson2023}, figure S7).
This behavior can be rationalized by recognizing that the TDEP effective force constant matrix $\bPhit$ is constructed to reproduce the exact \emph{static} correlation function, thus providing an accurate description of dynamical phonons only in the $\omega \rightarrow 0$ limit. While this approximation holds well for low-energy modes such as the M-tilt, its accuracy diminishes for higher-frequency phonons.
The seemingly improved agreement of the SCHA auxiliary phonons at high frequencies stems from a partial cancellation of the real part of the \emph{bubble} self-energy, similar to what is observed in the molecular case discussed in \figurename~\ref{fig:dynamical:spectral}.

This casts doubt on the spectral function reported in ref.\cite{LaniganAtkins2021} (figure 3, panel f) for the cubic phase of \ch{CsPbBr3} within the TDEP framework, where the acoustic mode becomes imaginary in the M–R region, and the resulting spectral function lacks the expected peak at real frequencies.

The error of not considering the full RPA resummation in the M-tilting mode is reflected in the dynamical spectral function, reported in \figurename~\ref{fig:CsSnI3}\textbf{(b)}. The spectrum exhibits a marked deviation from a Lorentzian lineshape, both when computed using the \emph{bubble} approximation and with the full RPA, reflecting the complex density-of-state accessible for phonon-phonon scattering events in solids compared to molecules (\figurename~\ref{fig:dynamical:spectral}). Here, the commonly employed bubble truncation, both in the SCHA, SCP, and in TDEP, fails to capture the prominent peak at \SI{6}{\per\centi\meter}, with the spectral weight instead shifted towards imaginary frequencies. This casts doubt on the spectral function reported in ref.\cite{LaniganAtkins2021} (figure 3, panel f) for the cubic phase of \ch{CsPbBr3} within TDEP, where the acoustic mode becomes imaginary in the M-R region and the resulting spectral function does not show any features. 
This instability likely originates from the overcorrection induced by the bubble diagram in the TDEP dynamical self-energy, as discussed in \secname~\ref{sec:dynamical}. Notably, the experimental data in that work do not exhibit a distinct low-frequency peak either, although it is plausible that such a feature lies below the experimental resolution. This difference, while it may seem subtle, plays a dominant role in the calculation of thermoelectric and thermodynamic properties, where quantities diverge as the phonon frequencies soften ($\omega\to 0$).

\section{Conclusions}
\label{sec:conclusions}

This work systematically compared different methodologies to simulate strongly anharmonic crystals. The first-principle grounds for the TDEP introduced here shed new light on the different TDEP implementations, providing a unique recipe to perform accurate simulations. To summarize them: i) start fitting only up to the second-order force constants, ii) employ a progressive fit over residual forces to extract higher-order contributions. iii) Always use the least-squares cost function on forces for fitting. iv) The quadratic term in the fit is the static-response function, without requiring any extra correction. v) Properly account for boundary conditions in Kramers-Kronig relations, and remove the self-energy for $\omega\to 0$ for any diagram computed (as reported in \eqname~\ref{eq:dyn:tdep:fixed}). Notably, (i-iv) are the default implementation of the original TDEP software\cite{knoop_tdep_2024} (but not of ref.\cite{bottin_-tdep_2020}), while (v) is not implemented yet in any available software to date. 
In particular, (v) is fundamental to recover the correct perturbative limit. 

We unveiled how SCHA and TDEP share a common origin as variational principles on the free energy and explained the origin of the observed discrepancies\cite{Fransson2023}. Both of them fail to reproduce the perturbative limit of the free energy. While it is possible to correct the SCHA and SCP a posteriori\cite{Oba_2019}, this violates the variational principle, which may compromise the accuracy of a phase diagram estimation.

Despite the failure of standard TDEP implementations in reproducing the perturbative limit of the phonon energies, reasonable results can still be obtained for the thermal transport properties, as the dominant contribution arises from the phonon lifetimes, which are correct in the perturbative limit regardless of the kind of TDEP implementation adopted. Indeed, this study reveals how much more care must be taken for strongly anharmonic crystals, where the real part of the phonon-phonon self-energy may play a crucial role and the \emph{bubble} approximation may fail, where the TD-SCHA is still the most reliable approximate method available\cite{monacelli_black_2020,Cherubini2021,Ranieri2024}, especially for correcting anharmonicity beyond the leading order.

\section*{Acknowledgments}
I acknowledge Giovanni Caldarelli, Antonio Siciliano, and Francesco Mauri for the helpful discussions we shared.

\appendix 
\section{Proof of the TDEP method}
\label{app:tdep:proof}

While introduced long ago, a rigorous formal derivation of the TDEP method is still difficult to find in the literature, as it is typically presented as a generic fit of the potential energy landscape on a finite temperature MD run. Consequently, the TDEP method comes with many flavors, such as the order of the temperature-dependent force constants and the cost function employed in the fit, usually chosen empirically. For this reason, here we derive the TDEP method from the variational principle of the free energy presented in the main text, namely \eqname~\eqref{eq:tdep:start}, which defines a unique and precise strategy in which the fit on the MD trajectory must be performed.
The $\bRcalt$ and $\bPhit$ parameters maximizing the free energy satisfy \eqname~\eqref{eq:tdep:start}.
For the centroids $\bRcalt$, the only term in the TDEP free energy depending on it is the auxiliary potential $\mathcal V_{\bRcalt,\bPhit}(\bm r)$, from which we get the condition
\begin{equation}
    \frac{\partial}{\partial \Rcalt_a} \sum_{ij}\left< (r_i - \Rcalt_i)\Phit_{ij}(r_j - \Rcalt_j)\right>_\text{MD} = 0,
\end{equation}
which trivially leads to
\begin{equation}
    \Rcalt_a = \left<r_a\right>_\text{MD},
\end{equation}
corresponding to \eqname~\eqref{eq:rc:tdep}. A bit more complex is the equation for $\bPhit$, as both $\mathcal V_{\bRcalt,\bPhit}(\bm r)$ and the harmonic free energy $F_{\bPhit}$ depend on it.
Let us start with the derivative of the auxiliary harmonic free energy, which can be computed from the partition function as
\begin{equation}
    F_\bPhit = -\frac{1}{\beta}\ln{\mathcal Z}_\bPhit.
    \label{eq:free:def}
\end{equation}
By replacing \eqname~\eqref{eq:part:func} into \eqname~\eqref{eq:free:def}, we get
\begin{equation}
    F_\bPhit = k_BT \left[\frac 12 \ln\left(\det\bPhit\right) - \sum_{i = 1}^{3N}\ln\left(2\pi k_BT \sqrt{m_i}\right)\right]. 
\end{equation}
Thus, the derivative of the free energy on $\bPhit$ only depends on the first term
\begin{equation}
    \frac{\partial F_\bPhit}{\partial \Phit_{ab}} = \frac{k_BT}{2} \frac{\partial}{\partial \Phit_{ab}} \ln\left(\det\bPhit\right)
\end{equation}
We can now exploit the definition of the determinant. Let $\lambda_k$ be the $k$-th eigenvalue of the $\bPhit$ symmetric matrix, and $e_k^i$ be the $i$-th component of the corresponding $k$-th eigenvector, the determinant is the product of the eigenvalues
\begin{align}
    \frac{\partial}{\partial \Phit_{ab}}\ln\left(\det\bPhit\right) &= \frac{\partial}{\partial \Phit_{ab}}
     \ln\left(\prod_{k = 1}^{3N}\lambda_k\right) \nonumber \\
    &= \frac{\partial}{\partial \Phit_{ab}}\sum_{k = 1}^{3N} \ln\lambda_k \nonumber \\
    & = \sum_{k=1}^{3N}\frac{1}{\lambda_k}\frac{\partial\lambda_k}{\partial \Phit_{ab}}
\end{align}
The derivative of the $k$-th eigenvalue on the Matrix element $\Phit_{ab}$ can be evaluated within perturbation theory
\begin{equation}
    \frac{\partial\lambda_k}{\partial \Phit_{ab}} = e_k^a e_k^b,
\end{equation}
\begin{equation}
    \frac{\partial}{\partial \Phit_{ab}}\ln\left(\det\bPhit\right) = 
    \sum_{k=1}^{3N} \frac{e_k^a e_k^b}{\lambda_k}.
    \label{eq:app:lambdainv}
\end{equation}
The right-hand side of \eqname~\eqref{eq:app:lambdainv} is the inverse of the $\bPhit$ matrix decomposed on the basis that diagonalizes $\bPhit$. 
\begin{equation}
    \frac{\partial}{\partial \Phit_{ab}}\ln\left(\det\bPhit\right) = 
    \left(\bPhit^{-1}\right)_{ab}
    \label{eq:app:phitinv}
\end{equation}
Thus, we get
\begin{equation}
    \frac{\partial F_\bPhit}{\partial \Phit_{ab}} = 
    \frac{k_BT}{2}\left(\bPhit^{-1}\right)_{ab}
\end{equation}
The other part of the gradient comes from the average of the auxiliary Harmonic potential
\begin{equation}
    \frac{\partial F_\bPhit}{\partial\Phit_{ab}} - \frac{\partial\braket{\mathcal V_{\bRcalt,\bPhit}}_\text{MD}}{\partial\Phit_{ab}} = 0
\end{equation}
From this we get
\begin{equation}
    \frac{k_BT}{2}\left(\bPhit^{-1}\right)_{ab} - \frac 12 \left<(r_a - \Rcalt_a)(r_b - \Rcalt_b)\right>_\text{MD} = 0,
\end{equation}
and consequently \eqname~\eqref{eq:phi:tdep}.

\subsection{The TDEP fit}
\label{app:tdep:proof:fit}

Here, the equivalence of the TDEP definition in \eqname~\eqref{eq:phi:tdep} with the fit procedure of the potential energy surface (\eqname~\ref{eq:tdep:fit}) is demonstrated.
The condition minimizing \eqname~\eqref{eq:tdep:fit} is
\begin{equation}
    \frac{\partial\mathcal L}{\partial \Rcalt_a} = -\sum_{bc}\Phi_{ab}\Phi_{bc}\left<r_c -\Rcal_c\right>_\text{MD} = 0,
\end{equation}
from which we get the \eqname~\eqref{eq:rc:tdep}:
\begin{equation}
    \left<r_a\right>_\text{MD} = \Rcalt_a.
\end{equation}
The equation for $\bPhit$ is a bit more involved:
\begin{equation}
     \frac{\partial\mathcal L}{\partial \Phit_{cd}} = 2\left<\left[f_c + \sum_b\Phit_{cb}(r_b - \Rcalt_b)\right](r_d - \Rcalt_d)\right>_\text{MD} = 0
\end{equation}
\begin{equation}
    \sum_b \Phit_{cb}\left<(r_b - \Rcalt_b)(r_d - \Rcalt_d)\right>_\text{MD} = - \left<f_c(r_d - \Rcalt_d)\right>_\text{MD}
\end{equation}
\begin{equation}
    \sum_b \Phit_{cb}\left<(r_b - \Rcalt_b)(r_d - \Rcalt_d)\right>_\text{MD} =  \int dr (r_d - \Rcalt_d) \frac{\partial V}{\partial r_c}\frac{e^{-\beta V}}{Z}.
\end{equation}
Integrating by parts the last term, we get
\begin{equation}
 -\left<f_c(r_d - \Rcalt_d)\right>_\text{MD} = 
 k_BT\delta_{cd}.
\end{equation}
This is a particular case of the generalized equipartition theorem, stating
$$
\left<\frac{\partial H}{\partial q_k}q_h\right>_\text{MD} = \delta_{hk}k_BT,
$$
where, $q_k$ and $q_h$ are two canonical variables, and $H$ is the real Hamiltonian of the system.
With this equality, we get
\begin{equation}
    \sum_b \Phit_{cb}\left<(r_b - \Rcalt_b)(r_d - \Rcalt_d)\right>_\text{MD} = k_BT\delta_{cd}
\end{equation}
Which is satisfied by the expression of $\bPhit$ that maximizes the TDEP free energy (\eqname~\ref{eq:phi:tdep}).

\section{Static fluctuation-dissipation theorem}
\label{app:fluctuation:dissipation}

Here, we report the proof of the textbook fluctuation-dissipation theorem in the static limit and discuss why it can be applied to TDEP but not to the SSCHA.

Consider a uniform force $\bm \xi$ perturbing the system. The new equilibrium state solves the modified Hamiltonian
\begin{equation}
H'(\bm \xi) = H - \sum_b\xi_b (r_b - \Rcalt_b).
\end{equation}
The new average position of the $a$-th atom is
\begin{equation}
    \braket{r_a}_{H'(\bm\xi)} = 
    \frac{\int d^{3N}r \,d^{3N}p e^{-\beta\left[ H - \sum_b \xi_b(r_b - \Rcalt_b)\right]}r_a}{\int d^{3N}r \,d^{3N}pe^{-\beta\left[ H- \sum_b \xi_b(r_b - \Rcalt_b)\right]}}
\end{equation}
The static response function is the derivative
of the new average position with respect to the
external force.
\begin{equation}
    \frac{\partial \braket{r_a - \Rcalt_a}_{H'(\bm\xi)}}{\partial \xi_b} = \beta \left<(r_b - \Rcalt_b)(r_a - \Rcalt_a)\right>_\text{MD}.
\end{equation}
This is no longer true if $H$ is self-consistent like in the SSCHA, as we have to add an extra term in the derivative due to the derivative of the unperturbed Hamiltonian with respect to the new perturbed equilibrium state. This extra term gives rise to the complex system of linear equations reported in \eqname~\eqref{eq:linear:response:rc} and \eqref{eq:linear:response:phi}, derived in \appendixname~\ref{app:linear:response:scha}.

\section{SCHA static linear response}

\label{app:linear:response:scha}

Here, the static linear response for the SSCHA theory is obtained. The derivation follows the derivation performed for the full dynamical linear response presented in ref.\cite{monacelli_time-dependent_2021} and coincides with the final result presented in ref.\cite{bianco_second-order_2017}.
Let us compute the derivative
\begin{equation}
    \frac{\partial\left< f_a\right>_{\Rcals,\Phis}}{\partial\xi_b} = \int d^{3N}{\bm R} \frac{\partial\rho_{\Rcals,\Phis}}{\partial\xi_b} f_a(\bR),
\end{equation}
where $\rho_{\Rcals,\Phis}(\bR)$ is the SCHA probability density of the SCHA.
The derivative of the density with respect to the Lagrange parameter $\xi_b$ can be obtained by using the chain rule on the two variational parameters of the SCHA: $\bRcals$ and $\bPhis$.

For brevity, the explicit indexing of $\bRcals$ and $\bPhis$ in $\rho$ is avoided in the rest of this appendix.
The derivative of the density with respect to
\begin{equation}
    \frac{\partial\rho}{\partial\xi_b} = \sum_{h} \frac{\partial\rho}{\partial\Rcals_h}\frac{\partial\Rcals_h}{\partial \xi_b} + 
    \sum_{hk}\frac{\partial\rho}{\partial \Phis_{hk}}\frac{\partial\Phis_{hk}}{\partial\xi_b},
\end{equation}
and
\begin{equation}
    \frac{\partial\rho}{\partial \Rcals_h} = -\rho(\bm R)\sum_k \Upsilon_{hk}(R_k - \Rcals_k),
\end{equation}
\begin{equation}
    \frac{\partial\rho}{\partial\Phis_{hk}} = \frac{\rho(\bR)}{2}\sum_{lm}\frac{\partial\Upsilon_{lm}}{\partial\Phis_{hk}}\left[\Upsilon_{lm}^{-1}- (R_l - \Rcals_l)(R_m - \Rcals_m) \right],
\end{equation}
where the $\bUps$ matrix is the inverse of the covariance matrix
\begin{equation}
    \left(\bUps^{-1}\right)_{ab} = \left<(R_a - \Rcals_a)(R_b - \Rcals_b)\right>,
\end{equation}
in the classical limit, it becomes:
\begin{equation}
    \Upsilon_{ab} = \frac{\Phi_{ab}}{k_BT},
\end{equation}
which is the classical counterpart of the quantum equation usually reported in the SSCHA papers\cite{bianco_second-order_2017,monacelli_stochastic_2021}:
\begin{equation}
\Upsilon_{ab} = \sqrt{m_am_b}\sum_\mu\frac{2\omega_\mu}{\hbar (2n_\mu + 1)} e_\mu^a e_\mu^b,
\end{equation}
where $\omega_\mu$ are SCHA frequencies (square root eigenvalues of the mass rescaled $\bPhis$ matrix), $e_\mu^a$ is the $a$-th Cartesian component of the $\mu$-the eigenvector, and $n_\mu$ the corresponding Bose-Einstein occupation factor.  
Exploiting these relationships, we get
\begin{align}
    \frac{\partial\left<f_a\right>}{\partial\xi_b}& =
    \sum_{hkl}\Upsilon_{kl}\left<(R_l - \Rcals_l)f_a\right>\frac{\partial \Rcals_h}{\partial \xi_b} + \nonumber \\
    &-\frac 12 \sum_{lm}\frac{\partial\Upsilon_{lm}}{\partial\Phis_{hk}}\left<(R_l - \Rcals_l)(R_m - \Rcals_m)f_a\right>\frac{\partial\Phis_{lm}}{\partial\xi_b}.
\end{align}
To obtain the other response function, we must exploit the self-consistent equation of the SCHA
\begin{equation}
    \Phis_{ab} = \left<\frac{\partial^2 V}{\partial R_a \partial R_b}\right>.
\end{equation}
We can directly integrate this by parts to get
\begin{equation}
    \Phis_{ab} = -\sum_h \Upsilon_{ah}\left<(R_h- \Rcals_h) f_b\right>,
\end{equation}
and differentiate.
Notably, the explicit derivative of $\Rcals$ in the average is zero, as it gives an average of forces that is zero at equilibrium. Thus, the only remaining derivatives are the one in $\bUps$ and the one in the distribution $\rho$
\begin{widetext}
\begin{align}
    \frac{\partial\Phis_{ab}}{\partial\xi_c}& = 
    -\sum_{hkl}\frac{\partial\Upsilon_{ah}}{\partial\Phis_{kl}}\frac{\partial\Phis_{kl}}{\partial\xi_c} \left<(R_h -\Rcals_h)f_b\right> + \nonumber \\ 
    & +\sum_{khl}\Upsilon_{ah}\Upsilon_{kl}\left<(R_h - \Rcals_h)(R_l - \Rcals_l)f_b\right>\frac{\partial\Rcals_k}{\partial\xi_b} + \nonumber \\
    & + \frac 12 \sum_{hklmn}\frac{\partial \Upsilon_{kl}}{\partial\Phis_{mn}}\Upsilon_{ah}\left<(R_h-\Rcals_h)(R_k-\Rcals_k)(R_l - \Rcals_l)f_b\right>\frac{\partial\Phis_{mn}}{\partial\xi_b} + \nonumber \\ 
    &-
    \frac 12 \sum_{hklmn}\frac{\partial\Upsilon_{kl}}{\partial\Phis_{mn}}\Upsilon^{-1}_{kl}\Upsilon_{ha}\left<(R_h - \Rcals_h)f_b\right>\frac{\partial\Phis_{mn}}{\partial\xi_b}.
\end{align}
\end{widetext}
We can remove the SCHA harmonic force from $\boldsymbol{f}$. This term is zero for all averages except the one multiplied by the displacement three times:
\begin{align}
    \left<u_a u_b u_c \left(-\sum_e \Phi_{de}u_e\right)\right> = -\sum_e \Phi_{de}&\bigg( 
    \Upsilon^{-1}_{ab}\Upsilon^{-1}_{ce} + \nonumber \\
    & +\Upsilon^{-1}_{ac}\Upsilon^{-1}_{be} + \nonumber \\ 
    & +\Upsilon^{-1}_{ae}\Upsilon^{-1}_{bc}
    \bigg),
\end{align}
where $u_a = R_a - \Rcals_a$.
Now, a bit of algebraic manipulations of matrices results in
\begin{equation}
    \frac{\partial\Upsilon_{ab}}{\partial\Phis_{lm}} =\sum_{hk} \frac{\partial\Upsilon^{-1}_{hk}}{\partial\Phis_{lm}}\frac{\partial\Upsilon_{ab}}{\partial\Upsilon^{-1}_{hk}},
\end{equation}
\begin{equation}
    \frac{\partial\Upsilon_{ah}}{\partial\Upsilon^{-1}_{pq}} = - \Upsilon_{ap}\Upsilon_{qh},
\end{equation}
\begin{equation}
    \frac{\partial\Upsilon_{ab}}{\partial\Phis_{lm}} =-\sum_{hk} \frac{\partial\Upsilon^{-1}_{hk}}{\partial\Phis_{lm}}\Upsilon_{ah}\Upsilon_{kb}.
\end{equation}
The last derivative is the so-called $\bLambda$ matrix, related to the two-phonon propagator introduced by Bianco et al. \cite{bianco_second-order_2017}.
\begin{equation}
    \frac{\partial\Upsilon_{ab}^{-1}}{\partial\Phis_{lm}} = 2\Lambda_{ablm},
    \label{eq:app:Lambda}
\end{equation}
and the three and four-phonon scattering vertices
\begin{equation}
    \Phithree_{abc} = \left<\frac{d^3V}{dR_adR_bdR_c}\right>
\end{equation}
\begin{equation}
    \Phifour_{abcd} = \left<\frac{d^4V}{dR_adR_bdR_cdR_d}\right>
\end{equation}

Coupling all together, we get the final system
\begin{equation}
    \delta_{ab} =  -\sum_{h}\Phis_{ah} \frac{\partial\Rcal_h}{\partial\xi_b}  + \sum_{pqlm}\Lambda_{pqlm}\Phithree_{pqa}\frac{\partial\Phis_{lm}}{\partial\xi_b} 
\end{equation}
\begin{equation}
    \frac{\partial\Phis_{ab}}{\partial\xi_c} =  -\sum_{h}\Phithree_{abh} \frac{\partial\Rcal_h}{\partial\xi_c}  -\sum_{pqlm}\Lambda_{pqlm}\Phifour_{pqab}\frac{\partial\Phis_{lm}}{\partial\xi_c} 
\end{equation}
\section{Perturbative expression of the free energy}
\label{app:free:diagrams}

Here, we derive the perturbative expression for the free energy and show that its diagrammatic representation coincides with the one reported in \figurename~\ref{fig:fe:diagrams}.

The expression of the free energy is given by
\begin{equation}
    F = -\frac{1}{\beta}\ln Z
\end{equation}
\begin{equation}
    Z = \int d^{3N}p e^{-\beta\frac{p^2}{2m}} \int d^{3N}R  e^{-\beta V(R)}
\end{equation}
Expanding perturbatively around a small third-order term, we get
\begin{equation}
    V(\bR) = \frac 1 2 \sum_{ab}\Phi_{ab}u_a u_b + \sum_{abc}\frac{1}{6}\Phithree_{abc}u_a u_b u_c
\end{equation}
where $\bm u$ represents the displacement from the average position.
Perturbatively, we can expand the exponential as
\begin{align}
    e^{-\beta V(R)} \approx &e^{-\frac\beta 2 \sum_{ab}\Phi_{ab}u_au_b} \bigg(
    1 - \frac\beta 6 \sum_{abc} \Phithree_{abc}u_a u_b u_c + \nonumber \\
    & \frac{\beta^2}{2\cdot 6^2}\sum_{\substack{abc\\def}}\Phithree_{abc}u_a u_b u_c u_d u_e u_f \Phithree_{def}\bigg)\label{eq:exp:pert}
\end{align}
Then we get
$$
Z = Z_0 + Z_1
$$
\begin{equation}
    \frac{Z_1}{Z_0} = \frac{\beta^2}{2\cdot 6^2}\sum_{\substack{abc\\def}}\Phithree_{abc}\left<u_a u_b u_c u_d u_e u_f\right>_0 \Phithree_{def}
    \label{eq:Z:pert}
\end{equation}
where $Z_0$ is the harmonic partition function, and the average $\left<\cdot\right>_0$ is performed on the harmonic Gaussian distribution. The odd term in \eqname~\eqref{eq:exp:pert} vanishes in when averaged over a Gaussian distribution as it is the integral of an odd function ($u^3$) times an even function (the Gaussian).
The full free energy is
\begin{equation}
    F \approx -\frac{1}{\beta}\ln\left[Z_0\left(1 + \frac{Z_1}{Z_0}\right)\right] = F_\text{harm} - \frac{1}{\beta}\frac{Z_1}{Z_0} 
\end{equation}
Therefore, the correction in \eqname~\eqref{eq:Z:pert} represents the perturbative correction beyond harmonic to the free energy, which is quadratic in the third-order term.
The six-body correlation function $\left<u_a u_b u_c u_d u_e u_f\right>_0$ can be evaluated by exploiting the properties of Gaussian integrals by splitting it in all the possible contractions of two-body correlations:
\begin{align}
    \left<u_a u_b u_c u_d u_e u_f\right>_0 =& 
    \left<u_a u_b\right>_0\left<u_c u_d\right>_0\left<u_e u_f\right>_0 + \nonumber \\
    & \left<u_a u_b\right>_0\left<u_c u_e\right>_0\left<u_d u_f\right>_0 + \nonumber \\
    & \left<u_a u_b\right>_0\left<u_c u_f\right>_0\left<u_d u_e\right>_0 + \cdots
    \label{eq:wick}
\end{align}
Each contraction is equal to the covariance matrix
\begin{equation}
    \left<u_au_b\right>_0 = \left(\bUps^{-1}\right)_{ab}.
\end{equation}
Exploiting the language from many-body theory, $\left<u_a u_b\right>_0$ is a static correlation function between two positions, i.e., the same time phononic Green function. In the language of Feynman diagrams, each one of the terms in \eqname~\eqref{eq:wick} can be represented by a line connecting two atoms $a$ and $b$, while the three-phonon vertex $\Phithree_{abc}$ as a triangle with the three atoms $a,b,c$ as vertices.
Therefore, the expression of the free energy correction in \eqname~\eqref{eq:Z:pert} can be written as two different terms:
\begin{equation}
\sum_{\substack{abc\\def}}\Phithree_{abc}\left<u_a u_b\right>_0\left<u_c u_d\right>_0\left<u_e u_f\right>_0 \Phithree_{def},
\label{eq:diag:loop}
\end{equation}
\begin{equation}
\sum_{\substack{abc\\def}}\Phithree_{abc}\left<u_a u_d\right>_0\left<u_b u_e\right>_0\left<u_c u_f\right>_0 \Phithree_{def}.
\label{eq:diag:bubble:three}
\end{equation}
All other terms in \eqname~\eqref{eq:wick} can be recast in one of \eqname~\eqref{eq:diag:loop} and \eqname~\eqref{eq:diag:bubble:three} by changing the names of the labels in the summation. \eqname~\eqref{eq:diag:loop} two indices of each three-phonons vertex are completely contracted with a phonon Green-function, leading to the Feynman diagram in \figurename~\ref{fig:fe:diagrams}\textbf{a}. \eqname~\eqref{eq:diag:bubble:three}, on the opposite, has all phonon Green's functions connecting different three-phonons vertices, and can be depicted as \figurename~\ref{fig:fe:diagrams}\textbf{b}.


%

\end{document}